\documentclass[journal=apchd5,manuscript=letter]{achemso}
\usepackage{amsmath,amssymb,graphicx,xcolor,braket,dcolumn,bm}
\usepackage[normalem]{ulem}

\newcommand{\Vb}{V_{\rm b}}
\newcommand{\Ib}{I_{\rm b}}
\newcommand{\Vg}{V_{\rm gate}}
\newcommand{\IV}{$I_{\rm b}$--$V_{\rm b}$}
\newcommand{\IonV}{I_{\rm b}(V_{\rm b})}
\newcommand{\dI}{dI_{\rm b}/dV_{\rm b}}
\newcommand{\dII}{d^2I_{\rm b}/dV^2_{\rm b}}
\newcommand{\mipt}{Center for Photonics and 2D Materials, Moscow Institute of Physics and Technology, Dolgoprudny, 141700, Russia}

\newcommand{\me}[1]{\textcolor{black}{#1}}

\usepackage{soul}

\title{Infrared photodetection in graphene-based heterostructures: bolometric and thermoelectric effects at the tunneling barrier}
 
\author{Dmitry A. Mylnikov$^1$}
\email{mylnikov.da@yandex.ru}
\affiliation{\mipt}

\author{Mikhail A. Kashchenko$^1$}
\affiliation{Programmable Functional Materials Lab, Center for Neurophysics and Neuromorphic Technologies, Moscow 127495}
\alsoaffiliation{\mipt}

\author{Kirill N. Kapralov}
\affiliation{\mipt}

\author{Davit A. Ghazaryan}
\affiliation{\mipt}
\alsoaffiliation{Laboratory of Advanced Functional Materials, Yerevan State University, Yerevan, 0025, Armenia}

\author{Evgenii E. Vdovin}
\affiliation{Institute of Microelectronics Technology RAS, Chernogolovka, 142432, Russia}

\author{Sergey V. Morozov}
\affiliation{Institute of Microelectronics Technology RAS, Chernogolovka, 142432, Russia}

\author{Kostya S. Novoselov}
\affiliation{Institute for Functional Intelligent
Materials, National University of Singapore, Singapore, 117575, Singapore}

\author{Denis A. Bandurin}
\email{bandurin.d@gmail.com}
\affiliation{Department of Materials Science and Engineering, National University of Singapore, 117575, Singapore}

\author{Alexander I. Chernov}
\affiliation{\mipt}
\alsoaffiliation{Russian Quantum Center, Skolkovo Innovation City, Moscow, 121205, Russia}

\author{Dmitry A. Svintsov}
\email{svintcov.da@yandex.ru}
\affiliation{\mipt}


\begin{document}

\begin{abstract}

Graphene/hBN/graphene tunnel devices offer promise as sensitive mid-infrared photodetectors but the microscopic origin underlying the photoresponse in them remains elusive. In this work, we investigated the photocurrent generation in graphene/ hBN/graphene tunnel structures with localized defect states under mid-IR illumination. We demonstrate that the photocurrent in these devices is proportional to the second derivative of the tunnel current with respect to the bias voltage, peaking during tunneling through the hBN impurity level. We revealed that the origin of the photocurrent generation lies in the change of the tunneling probability upon radiation-induced electron heating in graphene layers, in agreement with the theoretical model that we developed. Finally, we show that at a finite bias voltage, the photocurrent is proportional to the either of the graphene layers heating under the illumination, while at zero bias, it is proportional to the heating difference. Thus, the photocurrent in such devices can be used for accurate measurements of the electronic temperature providing a convenient alternative to Johnson noise thermometry.

\end{abstract}

\maketitle

\def\thefootnote{1}\footnotetext{These authors contributed equally to this work}\def\thefootnote{\arabic{footnote}}


\section{Introduction}




Mid-infrared (mid-IR) photodetectors hold immense importance across diverse fields. Capturing and visualizing thermal radiation they enable study of celestial objects and their evolution \cite{petric_mid-infrared_2011, rieke_mid-infrared_2015}, diagnostics and therapeutics through non-invasive imaging \cite{ring_infrared_2015}, nondestructive testing of components and detecting defects \cite{ciampa_recent_2018}. Mid-IR light carries information about molecular vibrations, providing valuable information about chemical composition of materials for monitoring environmental pollutants \cite{popa_towards_2019}.

Tunneling devices based on van der Waals heterostructures are attractive for infrared detection due to the presence of strong phonon-polariton absorption lines in many layered dielectrics~\cite{Polaritons_2DMs,PhPs_MoO3,PhPs_MoO3_2,PhPs_V2O5}, the most prominent example being the hyperbolic modes of hexagonal boron nitride (hBN)~\cite{PhPs_hBN,castilla_plasmonic_2020,PhPs_hBN2}. The intrinsic response time of tunneling detectors should be very short and limited only by vertical transport between few-nanometer thick barriers~\cite{massicotte_picosecond_2016,Fast_WG_integrated_tunnel_detector}. As the tunneling probability is exponentially sensitive to the electron energy and barrier height, the tunnel-type photodetectors may have additional photocurrent gain mechanisms. They can be associated either with carrier heating or light-induced charge accumulation in the layers. The latter effect is especially pronounced in multilayer structures such as those used in quantum well infrared photodetectors~\cite{QWIPs,QWIP_charge_buildup,Ryzhii_GLIPs}. Strong non-linearity of the tunneling current-voltage characteristic should give rise to a pronounced radiation rectification, at least at 'classical' electromagnetic frequencies when photon energy is below the barrier height~\cite{Liu_tunnel_nonlinearity,gayduchenko_tunnel_2021}. At 'quantum' frequencies, a more adequate picture of photocurrent is the photon-assisted tunneling~\cite{superlattice_Kazarinov}. The photon-aided tunneling should benefit from singularities in the joint density of states between tunnel-coupled layers. Such situation is realized in superlattice-based photodetectors~\cite{Rogalski2017,Quantum_cascade_detector} and quantum cascade lasers~\cite{QCL_Faist}, and is further anticipated for tunnel-coupled graphene layers~\cite{ryzhii_voltage-tunable_2014,Ryzhii_tunnel_injection_laser}. 

A variety of light-induced physical effects in tunnel-coupled 2D layers makes it challenging to reveal the dominant photodetection mechanism in such structures. The experimental studies of such structures were concentrated on the visible range, where both heating-induced photocurrents~\cite{xie_probing_2023} and direct photon-aided tunneling were detected~\cite{ma_tuning_2016}. A reverse process of tunneling accompanied by visible light emission was also observed in coupled graphene layers~\cite{kuzmina_resonant_2021}. The measurements of van der Waals tunneling photodetectors at lower electromagnetic frequencies are scarce. In the THz range, the measured photocurrents were speculated to originate from photon-aided tunneling~\cite{yadav_terahertz_2016}, yet no experimental proofs of this scenario were provided. As a result, the mechanism of photocurrent generation in graphene-based vertical tunnel structures in the infrared range remains unresolved.


In this work we report on the photocurrent measurements of graphene/hBN/graphene tunnel structure under IR illumination with a photon energy of 144--207~meV. The graphene layers in our structures are twisted by a large angle $\gtrsim 5$ degrees \cite{mishchenko_twist-controlled_2014,ghazaryan_twisted_2021}.
In such a situation, the direct current can be dominated by electron resonant hopping \text{via} impurity levels inside the hBN barrier,~\cite{chandni_signatures_2016, greenaway_tunnel_2018} where the \IV{}-characteristic becomes ladder-type with sharp slopes. We find that the photocurrent $I_{\rm ph}$ is proportional to the $\dII$ and is maximum under the condition of tunneling through the impurity level in hBN. At a non-zero bias, the photocurrent appears proportional to derivative of DC tunnel current with respect to the base cryostat temperature $d\Ib/dT$. The proportionality between these three quantities (the photocurrent, the curvature of \IV{}-characteristic, and the temperature derivative of DC current) is accurately reproduced by a theoretical model where the incident radiation causes electron heating in coupled graphene layers. At finite bias, the origin of photocurrent can be termed as bolometric effect across the tunnel junction. At zero bias, the photocurrent emerges only upon the asymmetric heating of electrons in the two layers. The phenomenon is analogous to the photo-thermoelectric effect, albeit it is developed across the tunnel barrier. We show that the knowledge of photocurrent at both zero and finite bias provides access to the electron temperature in individual layers. The proposed method is a simple alternative to the Johnson noise thermometry\cite{crossno_development_2015}, especially at cryogenic temperatures \cite{fong_ultrasensitive_2012, fong_measurement_2013, betz_supercollision_2013, tikhonov_noise_2016}.


\section{Results}
\subsection{Electrical characterization of tunneling infrared detector}
Our tunnel IR detector was fabricated using dry transfer technique~\cite{kretinin_electronic_2014} (see the Methods section for details). The structure consists of two flakes of single-layer graphene (SLG) separated by approximately three layers of hBN (thickness  $\sim 1$~nm). The structure is located on an hBN/graphite stack, the latter serving as a back gate. 
The intersection area of the top and bottom graphene is approximately 2.3~$\mu$m$^2$, which determines the area of the tunnel junction. A scheme of the stack and measurement configuration are presented in Figure 1a. A photograph of the heterostructure with marked contacts used for measurements is shown in Figure 1b. During measurements, the device was held in a cryostat at a base temperature of 7~K, unless otherwise is indicated.

\begin{figure*}
    \includegraphics[width=1\textwidth]{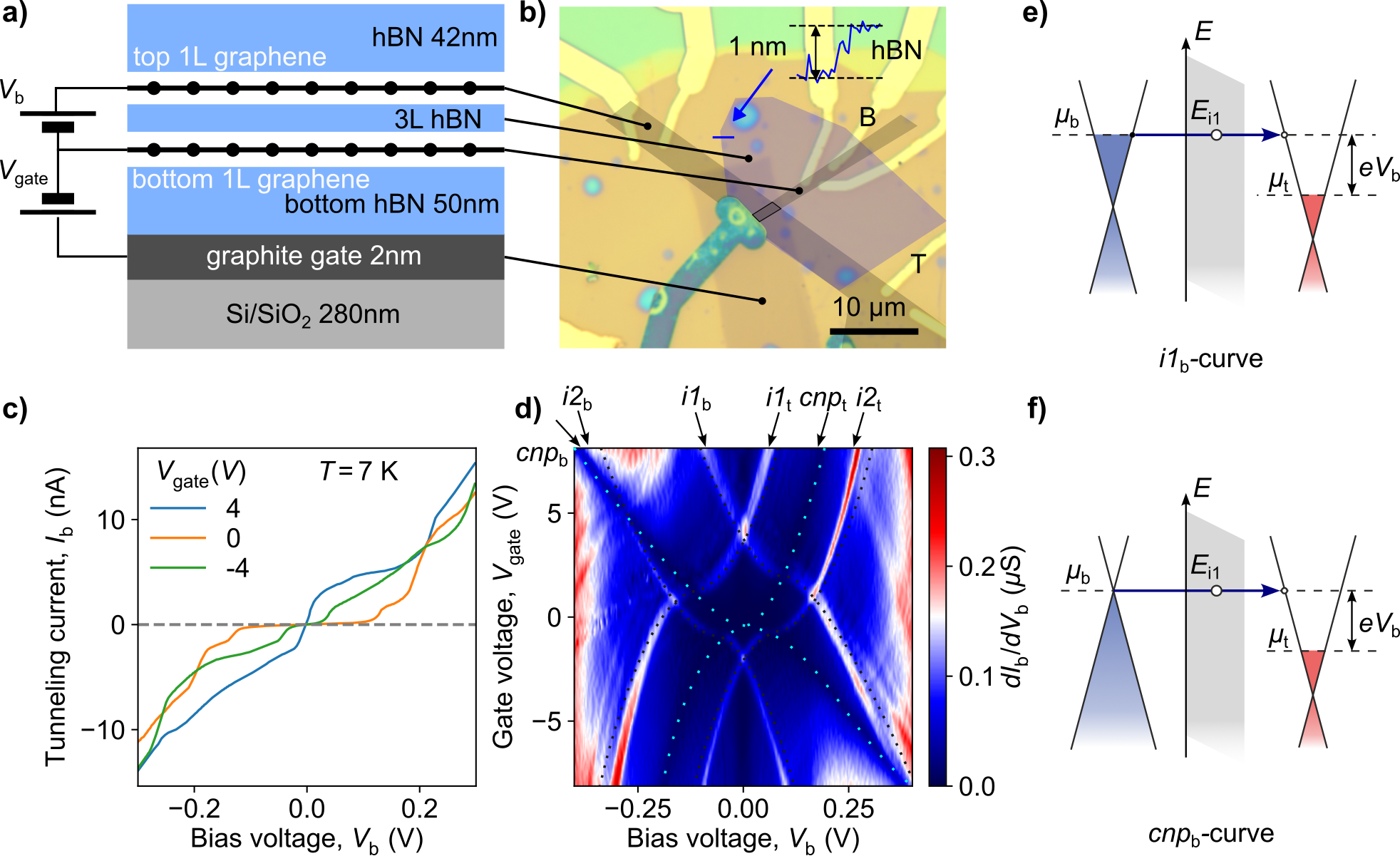}
    \caption{Tunneling detector scheme and electrical characterization.
    (a) The stack scheme of the tunneling detector. (b) Optical photograph of the sample with false-color images of the top, bottom graphene and barrier hBN superimposed on it. Contacts to the top and bottom graphene used in measurements are designated by the letters ``T'' and ``B'' accordingly. AFM scan of barrier hBN edge is shown in the inset. There is a cutout in the stack made to avoid possible shorting of the top and bottom graphene. (c) \IV{} device characteristics at different gate voltages, showing ladder-type behaviour. (d) Map of differential conductance as a function of bias and gate voltages at 7~K. The curves of maximum conductance are marked with $i1_{\rm t}$, $i2_{\rm t}$, $i1_{\rm b}$, $i2_{\rm b}$ and arrows,  of minimum conductance---$cnp_{\rm t}$, $cnp_{\rm b}$ and blue dotted lines. The black dotted lines correspond to theory. (e) Illustration of the tunneling process when the 1st impurity is aligned with the bottom graphene Fermi level, which corresponds to the $i1_{\rm b}$-curve on (d). (f) Same illustration but the impurity is aligned with the bottom graphene CNP ($cnp_{\rm b}$-curve on (d)). } 
    \label{fig1}
\end{figure*}

Electrical measurements of the tunnel structure (Figure 1c) confirm the rotational misalignment of graphene layers and the presence of resonant states in the dielectric. First of all, no traces of negative differential resistance are observed in the \IV{}-curves, which implies strong misorientation of graphene crystal structures. Second, the current-voltage characteristic has pronounced “steps”, which indicate the opening of new tunneling channels with increase in the bias voltage. The detailed mapping of differential conductance $\dI{}$ vs gate and bias voltages shown in Figure~1d confirms that these tunneling paths are associated with resonant passage of electrons through the impurity levels. We identify four characteristic spike lines in the differential conductance map, marked as $i1_{\rm t}$, $i2_{\rm t}$, $i1_{\rm b}$ and $i2_{\rm b}$. The spike positions depend on both bias and gate voltage, the latter controlling the carrier density. This excludes the phonon-assisted origin of conduction. Instead, each spike can be attributed to the alignment of the Fermi level in either graphene layer $\mu_{\rm t,b}$ with the level of impurity inside the band gap of boron nitride (Figure~1e. More precisely, the resonant condition can be formulated as
\begin{equation}
\label{Eq-resonant-condition}
    \mu_{\rm t,b}(\Vb,\Vg) = E_{i,n} + e F(\Vb,\Vg) x_{i,n},
\end{equation}
where $E_{i,n}$ is the energy level of $n$-th impurity in the absence of bias, the second term represents the bias-induced shift of impurity level, $F$ is the electric field in the hBN barrier, and $x_{i,n}$ is the position of the $n$-th impurity.


We manage to reproduce the experimentally measured positions of conduction spikes with the model (\ref{Eq-resonant-condition}) assuming two resonant levels within the barrier. The fitting procedure yields the defect levels $E_{i1}=100$ meV and $E_{i2}=-70$~meV reckoned from the Dirac points of unbiased layers. Further fitting enables the determination of impurity positions: the impurity \#1 is located in the middle of the barrier, while impurity \#2 is between the second and third hBN layers, counted from the bottom graphene. We also observe many other fainter resonant lines (curves near $i1$ and parallel to it) which emerge from other impurities, probably having smaller overlap with tunnel barrier. 

Another prominent feature of the conduction map is the area with nearly-zero differential conductivity. It forms two dark blue curves in Figure~1d, labeled $cnp_{\rm t}$ and $cnp_{\rm b}$. It corresponds the neutrality point of either graphene layer and, hence, to zero tunneling density of states (DOS) (Figure~1f). Both $cnp$-curves cross at nearly-zero gate and bias voltages, which implies the absence of initial doping in both graphene layers. It can be noted that tunneling through impurity does not occur on these curves, lines $i1$ and $i2$ are interrupted.



\begin{figure*}
    \includegraphics[width=1\textwidth]{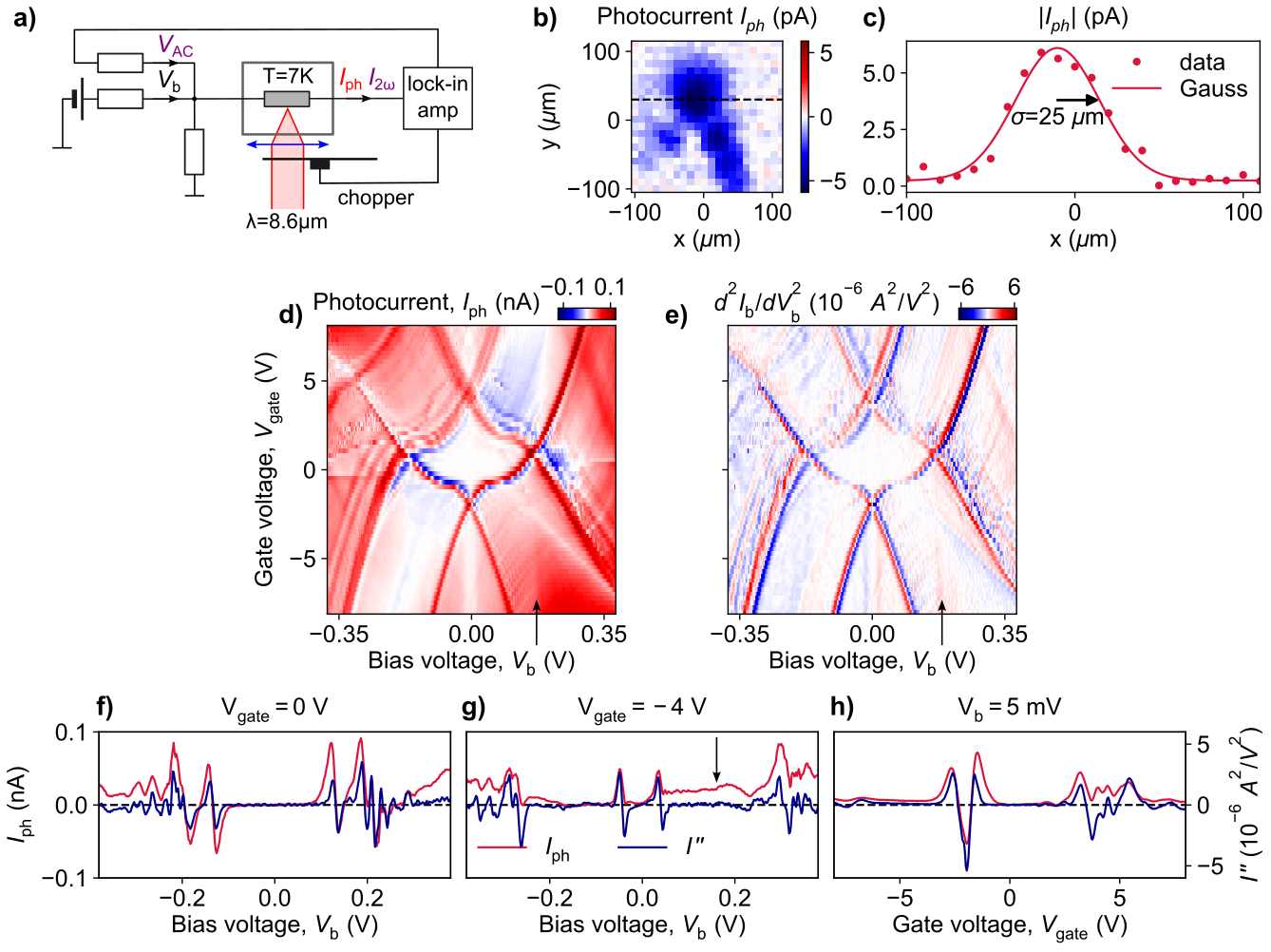}
    \caption{Photocurrent measurements under $\lambda=8.6$~$\mu$m illumination at $T=7$~K. (a) Scheme of the photocurrent and $\dII$ measurements. (b) Spatial photocurrent map. Deviation of the spot from the symmetrical Gaussian shape is aberration due to slightly oblique incidence of light on lens but not the sample features. (c) Slice of the map at $y=30$~$\mu$m and spot size extracted by fitting to Gaussian distribution. (d) Photocurrent as a function of bias and gate voltages. (e) Second derivative of $\IonV$ as a function of bias and gate voltages. Its clearly seen how the photocurrent repeats all $\dII$ features in details. (f)--(h) Slices of maps (d)--(e) at two different gate voltages: (f) $\Vg=0$~V, (g) $\Vg=-4$~V and (h) at small bias $\Vb=5$~mV. On (d),(e),(g) arrows show features at the energy of the optical phonon mode.}
    \label{fig2}
\end{figure*}


\subsection{Photocurrent under mid-infrared illumination}
We proceed to the characterization of the tunneling structure in the photodetector mode. The device is illuminated with mid-infrared radiation $\lambda=8.6$~$\mu$m fed from quantum cascade laser with output power \me{$P\sim 5$ mW}. The photocurrent is measured using lock-in amplifier (see Figure 2a and Methods for details). The photocurrent map recorded by moving the laser beam across the device represents a single bright spot (Figure 2b,c). This excludes the role of contact effects in photocurrent generation process, and shows that tunnel structure itself acts as a photocurrent generator.

Figure 2d shows the gate- and bias-resolved map of the photocurrent. First, the correlation of the photocurrent extrema with the position of the impurities on the differential conductance map is striking. The photocurrent extrema very closely follow the curves $i1$ and $i2$ of the differential conductance map, i.e. are observed when the Fermi levels of impurities and graphene layers are aligned. The next interesting feature is that when $\Vb$ changes and the impurity level passes through the Fermi level of one of the graphene layers, the photocurrent has two extrema of opposite signs, which are observed at $|\Vb|<0.25$~V, $|\Vg|<3$~V. At large values of the bias and gate, these features persists, yet the photocurrent acquires a positive 'background' growing with absolute value of bias.


The correlation between optoelectronic and electrical properties becomes even more apparent upon comparison of the photocurrent map with the $\dII$ map shown in Figure 2e. They match in the smallest details. The photocurrent extrema follows the $\dII$ extrema for all values of the bias and gate, repeating the position dependencies on $\Vg$ and $\Vb$ not only for the two main impurities, but also for all others. For example, with negative gate and offset values there are several impurities nearby. The photocurrent is amplified at each. At small values of $\Vb$ and $\Vg$ (namely, at $|\Vb|<0.25$~V, $|\Vg|<3$~V) there is a direct proportionality between the photocurrent and $\dII$ clearly visible on the map slices in Figure 2f-h. At larger gate and bias voltages, the presence of the background in photocurrent weakens this similarity, as indicated in Figure 2g. In this regime, $\dII$ has a strong sign-changing feature upon crossing the impurity level. The photocurrent does not change sign at these bias voltages, yet it demonstrates a spike at these points.


In addition, at $\Vb\approx180$--200~mV the photocurrent map shows features that do not depend on $\Vg$, expressed in a sudden increase in $I_{\rm ph}$ with increasing $\Vb$. This energy is greater than the energy of the incident photon (144~meV) and coincides with the energy of the optical phonon modes of graphene. On the map $\dII$ this feature is also visible as a local maximum. It is marked with a black arrows in Figures 2d,e,f. This coincides with previously observed phonon modes of graphene obtained from transport measurements \cite{chandni_signatures_2016, vdovin_phonon-assisted_2016}. Moreover, while low-energy phonon modes have already been observed as phonon assisted photocurrent under illumination with visible light, high-energy optical modes have been demonstrated.


All the key features of the dependencies of the photocurrent on the gate and bias, namely the proportionality of the photocurrent to the 2nd derivative of $\IonV$, the presence of two extrema of the opposite sign when the levels of graphene and impurity are aligned, were repeated in the measurements of the device \#2 (Supporting Information, Section II). The device was a graphene/hBN/graphene stack encapsulated in protective hBN layers with the same hBN barrier thickness of 1~nm. The stack was different in that it was made of multilayer graphene (2L on top and 3L on the bottom), had a large tunnel junction area, a silicon gate, and was illuminated at a different wavelength of $6.0$~$\mu$m.

\subsection{Origins of photocurrent: bolometric and thermoelectric effects at the tunnel barrier}
The two competing mechanisms contributing to the photocurrent in tunnel-coupled low-dimensional systems are the photon-assisted tunneling and the tunneling of hot carriers. The direct rectification by nonlinearity of the tunneling \IV{}-characteristic should be considered as a low-frequency limit of the photon-aided tunneling and does not require a separate consideration. Were the photon-aided tunneling a dominant light detection mechanisms, the photocurrent should peak when the Fermi level $\mu_{\rm t,b}$ plus the photon energy $\hbar\omega$ reaches the level of impurity $E_i$~\cite{fainberg2013photon, platero2004photon,kleinekathofer2006switching,tien1963multiphoton}. Such energy constraint corresponds to the onset of tunneling by photoexcited carriers along the resonant levels. The photocurrent map for this detection mechanism should posses extrema curves parallel to the impurity lines $i1_{\rm t}$, $i2_{\rm t}$ and shifted by $\hbar\omega = 144$ meV on the bias scale. We observe no peculiarities of $I_{\rm ph}(\Vb,\Vg)$ at these energies, and therefore exclude photon-assisted tunneling from relevant photodetection mechanisms.

An alternative to the photon-assisted tunneling realized in case of fast inter-carrier energy exchange is the tunneling of electrons heated up by absorbed radiation. The light-induced change in electron temperature leads to the broadening of their Fermi distributions and, hence to the change in average tunneling probability. Further physics depends essentially on whether the layers are biased or not, and on the relative position of Fermi levels and impurity level.

For biased structures, the photocurrent represents the change in total tunneling current induced by carrier heating. If the Fermi level of a particular graphene layer is biased slightly below the impurity level {\it in the dark}, the light-induced heating would push the electrons toward the resonant level. This would increase the total current and lead to a positive spike in the photocurrent, (Figure~3a). If the same layer is biased slightly above the impurity level {\it in the dark}, the light-induced heating would deplete the Fermi distribution in the vicinity of resonant energy. In such situation, the negative spike in the photocurrent would be observed (Figure~3b). Such a double-spike structure of photocurrent is indeed observed each time the Fermi level $\mu_{\rm t,b}$ crosses the impurity level at not very large bias voltages (Figure 2f). This fact can already be considered as a proof of hot-carrier origin of the photocurrent.

At near-zero bias, the Fermi levels of both graphene layers are very close. When the impurity level is aligned with them the 2 effects described above are superimposed on each other, resulting in a picture with 3 spikes which we observed at $\Vb=5$~mV (Figure 2h).

At zero bias, the origin of photocurrent is distinct and can be termed as Seebeck effect across the tunnel barrier. Heating of both top and bottom electronic subsystems by the same amount cannot result in any current due to their partial thermal equilibrium. However, asymmetric heating of electrons in the layers would result imbalance between tunneling currents, measured as photocurrent. The photocurrent would be therefore proportional to the temperature difference between the layers, $I_{\rm ph} \propto T_{\rm t} - T_{\rm b}$.

\begin{figure*}
    \includegraphics[width=1\textwidth]{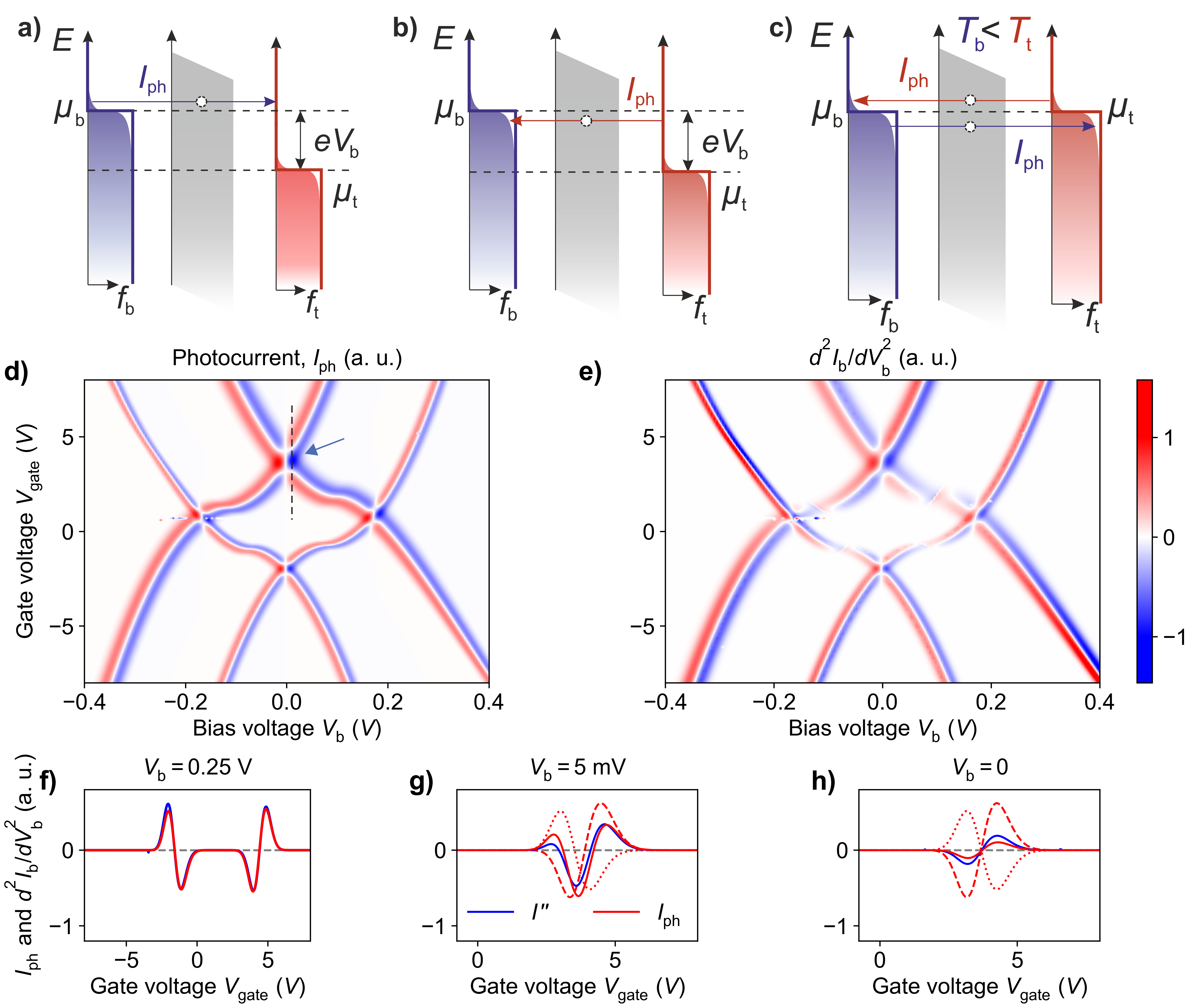}
    \caption{Illustration and theory of the photocurrent generation. (a)--(b) Illustration of the thermal mechanism of the photocurrent generation when the top graphene layer is biased slightly (a) below or (b) above the impurity level (which corresponds to $i1_{\rm b}$, $i2_{\rm b}$-curves on Figure 1d). (c) Illustration of the photocurrent generation at zero bias at 2 gate voltages near impurity alignment. (d) Theoretically calculated photocurrent and (e) $\dII$ map as a function of gate and bias voltages, well reproducing experimental results. (f)--(g) Slices of maps (d)--(e) at three different bias voltages: (f) $\Vb=0.25$~V; (g) $\Vb=5$~mV, along dashed line on (d); (h) $\Vb = 0$. The red dashed and dotted lines on (g),(h) demonstrate the contribution of the top and bottom layers, respectively, to the total photocurrent. The heating of the layers is assumed to be slightly different, $\delta T_{\rm t} = 1.2\delta T_{\rm b}$, which gives a non-zero photocurrent at $\Vb = 0$.
    }
    \label{fig3}
\end{figure*}

Further proofs of the thermal origin of the photocurrent can be obtained by direct calculation of temperature-dependent current-voltage characteristics $\Ib(T_{\mathrm{t}},T_{\mathrm{b}})$. We have obtained the latter with Bardeen transfer Hamiltonian approach
\begin{equation}
\label{Eq-current-general}
\Ib(T_{\mathrm{t}},T_{\mathrm{b}})= \frac{e}{\hbar} \int\limits_{-\infty}^{\infty} d E \left[f_{\mathrm{t}}(E)-f_{\mathrm{b}}(E)\right]{\mathcal{D}(E)},
\end{equation}
where $f_{\mathrm{t,b}}=\left[1+\exp \left(\left(E-\mu_{\mathrm{t,b}}\right) / k_{\mathrm{B}} T_{\mathrm{t,b}}\right)\right]^{-1}$ are the Fermi distribution functions in top and bottom layers with generally different temperatures ($T_{\mathrm{t}}$ and $T_{\mathrm{b}}$) and Fermi levels. The function $\mathcal{D}(E)$ is the energy-dependent tunneling probability possessing sharp resonances at impurity levels $E = E_{i,n}$ (see \cite{greenaway_tunnel_2018} and the Supporting Information, Sec. III for explicit form). The model (\ref{Eq-current-general}) is suitable both for calculations of DC source-drain current (with all its derivatives) and the photocurrent. For DC current, one sets the temperatures of both layers to the base cryostat temperature $T_{\mathrm{t}} = T_{\mathrm{b}} = T_0$. Evaluating the photocurrent, one sets $T_{\mathrm{t,b}} = T_0 +\delta T_{\mathrm{t,b}}$, where $\delta T_{\mathrm{t,b}}$ are the light-induced overheating of top and bottom layer. In explicit form
\begin{equation}
\label{Eq-photocurrent}
I_{\rm ph} = I(T_0 + \delta T_{\rm t},T_0 + \delta T_{\rm b}) - I (T_0). 
\end{equation}

The results of photocurrent calculations are presented in Figure 3d-h and fully confirm the above intuitive picture on light-induced heating effects. Under finite bias $\Vb$, the theoretically predicted photocurrent indeed possesses upward and downward spikes at both sides of impurity levels. Moreover, the theory reproduces the observed proportionality between photocurrent and $\dII$, which can be derived analytically in a fashion similar to the derivation of Wiedemann-Franz law  (see Supporting information, Sec. III for details).

At a bias close to zero, the photocurrent is a superposition of the photocurrent profiles from two graphene layers. At a bias exactly zero, their sum gives the dependence of the photocurrent on the gate in the form of two spikes of opposite signs, the amplitudes of which are proportional to the temperature difference of the graphene layers $I_{\rm ph} \propto \delta T_\mathrm{t} - \delta T_\mathrm{b}$ (Figure 3g).

However, applying even a small bias $\Vb=5$~mV spoils this ideal picture. The photocurrent profiles from the two layers are summed up into a pattern with 3 spikes, two upward (downward) spikes and a single downward (upward). A strong central spike mostly corresponds to the sum of the co-directional photocurrents from both graphene layers and represents the average heating of the graphene layers. While the two side spikes represent the sum of the opposing photocurrents from the two layers and contain information about the temperature difference between the layers (Figure 3h). This is exactly what we see in the experiment at $\Vb\approx0$~mV (Figure 2h). The accuracy of setting and measuring the voltage did not allow us to clearly catch the case of zero bias (Figure S3).


Once the origin of photocurrent is thermal, its functional dependence $I_{\rm ph}(\Vb,\Vg)$ should be similar to that of temperature conductance coefficient $d\Ib/dT_0$. Independent measurements of photocurrent and tunnel current $\Ib (T_0)$ with variable base cryostat temperature confirm this idea, as shown in Figure 4a,b.

More precisely, in a linear approximation of (3) on temperature, the photocurrent is given by
\begin{equation}
    I_{\rm ph} = \dfrac{dI_{\rm b}}{dT_\mathrm{t}} \delta T_\mathrm{t} + \dfrac{dI_{\rm b}}{dT_\mathrm{b}} \delta T_\mathrm{b}.
\end{equation}

When $e\Vb \gg k_{\rm B} T_0\approx0.7$~meV and away from the $i1_{\rm t}$, $i1_{\rm b}$, $i2_{\rm t}$, $i2_{\rm b}$ lines crossing areas determined by the energy width of the impurity levels ($\approx4$~meV), photocurrent depends only on $d\Ib/dT\delta T$ of one of the graphene layers, which Fermi level is aligned with the impurity level. The Fermi level of the other graphene layer is far away and does not significantly contribute to the photocurrent. It is become possible to calculate the increase in the electron temperature of each layer by dividing the measured photocurrent value by $d\Ib/dT$ at curves $i1_{\rm t}$, $i2_{\rm t}$ for the top layer and $i1_{\rm b}$, $i2_{\rm b}$ for the bottom.

We calculated the derivative of the current with respect to temperature as $d\Ib/dT=(\Ib(T_2)-\Ib(T_1))/(T_2-T_1)$, where $T_1$, $T_2$ are 10 and 20~K for Figure 4a, and 7 and 20~K for Figure 4b. Comparing its profile with the photocurrent measured at 7~K at the positions of spikes $i1_{\rm t}$, $i2_{\rm t}$, $i1_{\rm b}$, $i2_{\rm b}$ we estimate the average heating of electrons in both graphene layers at our incident light power \me{5~mW} to be $\delta T\approx8$~K. The spread of the top and bottom layer heating values does not allow us to estimate their difference.

In the unbiased case, we did not observe ideal photocurrent profiles with two spikes, only a pattern corresponding to a near-zero bias. This still allows us to evaluate the difference in heating of the graphene layers. Fitting calculated $I_{\rm ph}(\Vg)$ curves at $\Vb=5$~mV to the experimental data, we estimate the temperature difference between the layers to be $\lesssim 1.5$~K.

\begin{figure*}
    \includegraphics[width=1\textwidth]{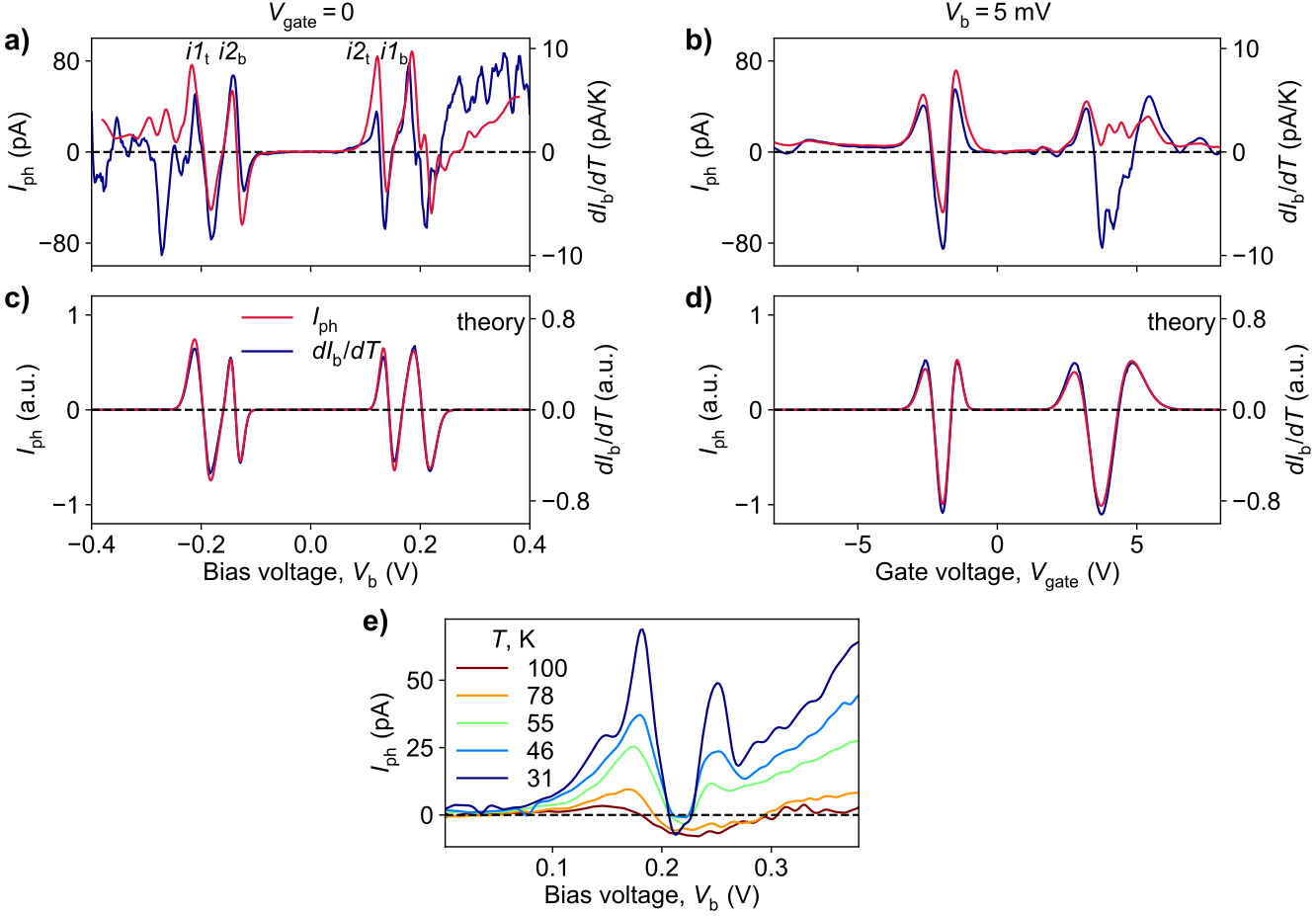}
    \caption{Estimation of the temperature of the electron gas heated by radiation from $d\Ib/dT$ and photocurrent measurements. (a) $d\Ib/dT$ and photocurrent as function of bias voltage measured at $\Vg=0$, (b) as function of gate voltage at small bias $\Vb=5$~mV. Electron temperature rise is estimated to be about 8~K. (c)--(d) Theoretically calculated photocurrent and $d\Ib/dT$ for (c) $\Vg=0$ and (d) $\Vb=5$~mV for $\delta T_{\rm t} = 1.2\delta T_{\rm b}$. (e) Photocurrent dependence on $\Vb$ at elevated temperatures.
    }
    \label{fig4}
\end{figure*}

As the temperature increases, the Fermi distribution blurs, which, according to the theory, leads to a decrease in $d\Ib/dT$ inversely proportional to temperature, and hence to a decreasing of the photocurrent. In Figure 4e the photocurrent dependence on $\Vb$ is presented at different temperatures from 31 to 100~K demonstrating a decline with increasing the temperature, as expected. 

\section{Discussion}

In summary, we have elucidated the mechanism of photocurrent generation in graphene/hBN/graphene tunnel structures with localized defect states in a barrier when illuminated with mid-IR light. Heating electrons by radiation broadens the Fermi distribution. In biased case the photocurrent arises due to a change in the probability of tunneling from one layer to another and occurs when the Fermi level of graphene and the impurity level are aligned. This is a thermal effect, thus the photocurrent is proportional to the heating of the electrons. At strictly zero bias, we have a different case and the photocurrent is already proportional to the temperature difference between the layers. However, when a few mV bias is applied, this picture blurs and becomes closer to the biased case.

We calculated the photocurrent as the change in the tunneling probability of light-heated electrons and reproduced the experimental data in detail at all bias values.

Measuring the thermal photocurrent allows one to calculate the temperature of electrons heated by radiation. Such measurements are simpler than Johnson noise thermometry, which is especially difficult at low temperatures. Measuring the heating of electrons at different lattice temperatures will allow us to calculate cooling rates and clarify the mechanisms of electron scattering.

The likely reason that we do not observe the photon-assisted mechanism is that the electron cooling rate is higher than the tunneling rate. Indeed, $\tau_{\rm el}=\varepsilon_F \hbar/(\hbar\omega)^2$ is on the order of fs, while $\tau_{\rm tun}= \hbar/E_{\rm bar}\cdot \exp(\sqrt{2m^*E_{\rm bar}} d_{\rm bar}/\hbar)$ is on the order of nanoseconds, where $E_{\rm bar}$ is height and $d_{\rm bar}$ is the thickness of hBN barrier. But the photon-assisted mechanism can become dominant at radiation frequencies $>3$~THz.

It is worth noting that the magnitude of the photocurrent can be significantly increased with a large area of the tunnel region and a small barrier thickness, as well as using WS$_2$ as a barrier layer \cite{georgiou_vertical_2013,bai_highly_2022}. On the presented device, the photocurrent was 120~pA. Device \#2, with 10 times greater conductivity, demonstrated 100 times greater photocurrent, up to 20~nA (Supporting Information, Section II).

Such tunnel micro-detectors demonstrating high photocurrent could be envisioned as a building block for multipixel mid-IR cameras.

\section{Materials and Methods}
\subsection{Device Fabrication}
Devices were made using dry transfer technique~\cite{kretinin_electronic_2014}. This involved standard dry-peel technique to obtain graphene and hBN crystals. The flakes were stacked on top of each other (from top hBN to bottom graphite) using a stamp made of PolyBisphenol carbonate (PC) on polydimethylsiloxane (PDMS) and deposited on top of an oxidized (280 nm of SiO$_2$) high-conductivity silicon wafer (KDB-0.001, $\sim$0.001--0.005~$\Omega\cdot$cm). The resulting  thickness of the hBN layers was measured by atomic force microscopy. Then electron-beam lithography and reactive ion etching with SF6  (30 sccm, 125 Watt power) were employed to define contact regions in the obtained hBN/graphene/barrier hBN/graphene/hBN/graphite heterostructure. Metal contacts were made by electron-beam evaporating 3~nm of Ti and 70~nm of Au. The second lithography was done to make a cutout to avoid possible shorting of the top and bottom graphene due to the displacement of the thin hBN layer during transfer. It was followed by reactive ion etching using PMMA as the etching mask.

\subsection{Measurements}
The sample was held at 7~K inside a cold finger closed-cycle cryostat (Montana Instruments, s50). \IV{} characteristics was measured using Keithley 2636B sourcemeter. Differential conductance was calculated by numerical $\IonV$ differentiating. $\dII$ was measured using AC-DC mixing technique. Source-meter (Keithley Instruments, 2636B) DC voltage and lock-in amplifier (Stanford Research, SR860) output AC sine voltage at 4~Hz frequency was summed up by voltage divider resulting in DC bias with small $V_{\rm AC}=4.8$~mV (rms) component applied to the sample. Second derivative of \IV{} was calculated from lock-in second harmonic readings $I_{2\omega}$ at +90° phase  as $\dII = -\sqrt{2} I_{2\omega} / {V_{\rm AC}^2}$. Photocurrent was also measured using lock-in amplifier. Linearly polarized light from a quantum cascade laser with a wavelength of 8.6~$\mu$m was modulated by a chopper at a frequency of 8 Hz. Light was focused by ZnSe lens through polypropylene film cryostat window to an almost diffraction limited spot. Motorized XY stage allowed precise aligning of sample and laser spot. Binding of chopper phase was done by comparing photocurrent phase with that obtained in case of laser current modulation and additionally controlled from amplified waveforms on the oscilloscope. Deeper details are presented in Supporting Information, Section I.

\begin{suppinfo}
(I) Measurements details; (II) Photocurrent measurements of the device \#2; (III) Theory of photocurrent generation.
\end{suppinfo}

\section*{Author contributions}
D.A.S., D.A.M. and D.A.B. designed and supervised the project; M.A.K., D.A.B. and K.S.N. fabricated the devices; D.A.M. performed the measurements and analyzed the experimental data with the help from D.A.S., D.A.B., D.A.G., A.I.C., S.V.M., and E.E.V.; K.N.K. and D.A.S. developed the theoretical model; D.A.M., K.N.K, and D.A.S. wrote the text with inputs from all authors. All authors contributed to the discussions.

\section*{Acknowledgments}
The work of D.A.M., M.A.K., K.N.K. and D.A.S. (photocurrent measurements and theoretical modelling) was supported by the Russian Science Foundation, grant \# 21-79-20225. M.A.K. acknowledges the support of an internal grant program at the Center for Neurophysics and Neuromorphic Technologies. The devices were fabricated using the equipment of the Center of Shared Research Facilities (MIPT). D.A.B. acknowledges the support of A*STAR YIRG grant M22K3c0106. K.S.N. is grateful to the Ministry of Education, Singapore (Research Centre of Excellence award to the Institute for Functional Intelligent Materials, I-FIM, project No. EDUNC-33-18-279-V12) and to the Royal Society (UK, grant number RSRP\textbackslash R\textbackslash190000) for support. E.E.V. and S.V.M. were supported by
the Russian Ministry of Science and Higher Education (grant \# 075-01304-23-00).

\section*{Competing interests}
All authors declare no financial or non-financial competing interests.

\bibliography{bibtexa,Bibliography-extra}

\providecommand{\latin}[1]{#1}
\makeatletter
\providecommand{\doi}
  {\begingroup\let\do\@makeother\dospecials
  \catcode`\{=1 \catcode`\}=2 \doi@aux}
\providecommand{\doi@aux}[1]{\endgroup\texttt{#1}}
\makeatother
\providecommand*\mcitethebibliography{\thebibliography}
\csname @ifundefined\endcsname{endmcitethebibliography}  {\let\endmcitethebibliography\endthebibliography}{}
\begin{mcitethebibliography}{47}
\providecommand*\natexlab[1]{#1}
\providecommand*\mciteSetBstSublistMode[1]{}
\providecommand*\mciteSetBstMaxWidthForm[2]{}
\providecommand*\mciteBstWouldAddEndPuncttrue
  {\def\EndOfBibitem{\unskip.}}
\providecommand*\mciteBstWouldAddEndPunctfalse
  {\let\EndOfBibitem\relax}
\providecommand*\mciteSetBstMidEndSepPunct[3]{}
\providecommand*\mciteSetBstSublistLabelBeginEnd[3]{}
\providecommand*\EndOfBibitem{}
\mciteSetBstSublistMode{f}
\mciteSetBstMaxWidthForm{subitem}{(\alph{mcitesubitemcount})}
\mciteSetBstSublistLabelBeginEnd
  {\mcitemaxwidthsubitemform\space}
  {\relax}
  {\relax}

\bibitem[Petric \latin{et~al.}(2011)Petric, Armus, Howell, Chan, Mazzarella, Evans, Surace, Sanders, Appleton, Charmandaris, Díaz-Santos, Frayer, Haan, Inami, Iwasawa, Kim, Madore, Marshall, Spoon, Stierwalt, Sturm, U, Vavilkin, and Veilleux]{petric_mid-infrared_2011}
Petric,~A.~O. \latin{et~al.}  Mid-{Infrared} {Spectral} {Diagnostics} of {Luminous} infrared {Galaxies}. \emph{The Astrophysical Journal} \textbf{2011}, \emph{730}, 28, Publisher: The American Astronomical Society\relax
\mciteBstWouldAddEndPuncttrue
\mciteSetBstMidEndSepPunct{\mcitedefaultmidpunct}
{\mcitedefaultendpunct}{\mcitedefaultseppunct}\relax
\EndOfBibitem
\bibitem[Rieke \latin{et~al.}(2015)Rieke, Wright, Böker, Bouwman, Colina, Glasse, Gordon, Greene, Güdel, Henning, Justtanont, Lagage, Meixner, Nørgaard-Nielsen, Ray, Ressler, Dishoeck, and Waelkens]{rieke_mid-infrared_2015}
Rieke,~G.~H. \latin{et~al.}  The {Mid}-{Infrared} {Instrument} for the {James} {Webb} {Space} {Telescope}, {I}: {Introduction}. \emph{Publications of the Astronomical Society of the Pacific} \textbf{2015}, \emph{127}, 584, Publisher: University of Chicago Press\relax
\mciteBstWouldAddEndPuncttrue
\mciteSetBstMidEndSepPunct{\mcitedefaultmidpunct}
{\mcitedefaultendpunct}{\mcitedefaultseppunct}\relax
\EndOfBibitem
\bibitem[Ring \latin{et~al.}(2015)Ring, Jung, and Żuber]{ring_infrared_2015}
Ring,~F.; Jung,~A.; Żuber,~J. \emph{Infrared {Imaging}: {A} casebook in clinical medicine}; IOP Publishing, 2015\relax
\mciteBstWouldAddEndPuncttrue
\mciteSetBstMidEndSepPunct{\mcitedefaultmidpunct}
{\mcitedefaultendpunct}{\mcitedefaultseppunct}\relax
\EndOfBibitem
\bibitem[Ciampa \latin{et~al.}(2018)Ciampa, Mahmoodi, Pinto, and Meo]{ciampa_recent_2018}
Ciampa,~F.; Mahmoodi,~P.; Pinto,~F.; Meo,~M. Recent {Advances} in {Active} {Infrared} {Thermography} for {Non}-{Destructive} {Testing} of {Aerospace} {Components}. \emph{Sensors} \textbf{2018}, \emph{18}, 609, Number: 2 Publisher: Multidisciplinary Digital Publishing Institute\relax
\mciteBstWouldAddEndPuncttrue
\mciteSetBstMidEndSepPunct{\mcitedefaultmidpunct}
{\mcitedefaultendpunct}{\mcitedefaultseppunct}\relax
\EndOfBibitem
\bibitem[Popa and Udrea(2019)Popa, and Udrea]{popa_towards_2019}
Popa,~D.; Udrea,~F. Towards {Integrated} {Mid}-{Infrared} {Gas} {Sensors}. \emph{Sensors} \textbf{2019}, \emph{19}, 2076, Number: 9 Publisher: Multidisciplinary Digital Publishing Institute\relax
\mciteBstWouldAddEndPuncttrue
\mciteSetBstMidEndSepPunct{\mcitedefaultmidpunct}
{\mcitedefaultendpunct}{\mcitedefaultseppunct}\relax
\EndOfBibitem
\bibitem[Low \latin{et~al.}(2017)Low, Chaves, Caldwell, Kumar, Fang, Avouris, Heinz, Guinea, Martin-Moreno, and Koppens]{Polaritons_2DMs}
Low,~T.; Chaves,~A.; Caldwell,~J.~D.; Kumar,~A.; Fang,~N.~X.; Avouris,~P.; Heinz,~T.~F.; Guinea,~F.; Martin-Moreno,~L.; Koppens,~F. {Polaritons in layered two-dimensional materials}. \emph{Nature Materials} \textbf{2017}, \emph{16}, 182--194\relax
\mciteBstWouldAddEndPuncttrue
\mciteSetBstMidEndSepPunct{\mcitedefaultmidpunct}
{\mcitedefaultendpunct}{\mcitedefaultseppunct}\relax
\EndOfBibitem
\bibitem[Zheng \latin{et~al.}(2018)Zheng, Chen, Wang, Wang, Chen, Liu, Xu, Xie, Chen, Deng, and Xu]{PhPs_MoO3}
Zheng,~Z.; Chen,~J.; Wang,~Y.; Wang,~X.; Chen,~X.; Liu,~P.; Xu,~J.; Xie,~W.; Chen,~H.; Deng,~S.; Xu,~N. Highly Confined and Tunable Hyperbolic Phonon Polaritons in Van Der Waals Semiconducting Transition Metal Oxides. \emph{Advanced Materials} \textbf{2018}, \emph{30}, 1705318\relax
\mciteBstWouldAddEndPuncttrue
\mciteSetBstMidEndSepPunct{\mcitedefaultmidpunct}
{\mcitedefaultendpunct}{\mcitedefaultseppunct}\relax
\EndOfBibitem
\bibitem[Ma \latin{et~al.}(2018)Ma, Alonso-Gonz{\'{a}}lez, Li, Nikitin, Yuan, Mart{\'{i}}n-S{\'{a}}nchez, Taboada-Guti{\'{e}}rrez, Amenabar, Li, V{\'{e}}lez, Tollan, Dai, Zhang, Sriram, Kalantar-Zadeh, Lee, Hillenbrand, and Bao]{PhPs_MoO3_2}
Ma,~W. \latin{et~al.}  {In-plane anisotropic and ultra-low-loss polaritons in a natural van der Waals crystal}. \emph{Nature} \textbf{2018}, \emph{562}, 557--562\relax
\mciteBstWouldAddEndPuncttrue
\mciteSetBstMidEndSepPunct{\mcitedefaultmidpunct}
{\mcitedefaultendpunct}{\mcitedefaultseppunct}\relax
\EndOfBibitem
\bibitem[Taboada-Guti{\'{e}}rrez \latin{et~al.}(2020)Taboada-Guti{\'{e}}rrez, {\'{A}}lvarez-P{\'{e}}rez, Duan, Ma, Crowley, Prieto, Bylinkin, Autore, Volkova, Kimura, Kimura, Berger, Li, Bao, Gao, Errea, Nikitin, Hillenbrand, Mart{\'{i}}n-S{\'{a}}nchez, and Alonso-Gonz{\'{a}}lez]{PhPs_V2O5}
Taboada-Guti{\'{e}}rrez,~J. \latin{et~al.}  {Broad spectral tuning of ultra-low-loss polaritons in a van der Waals crystal by intercalation}. \emph{Nature Materials} \textbf{2020}, \emph{19}, 964--968\relax
\mciteBstWouldAddEndPuncttrue
\mciteSetBstMidEndSepPunct{\mcitedefaultmidpunct}
{\mcitedefaultendpunct}{\mcitedefaultseppunct}\relax
\EndOfBibitem
\bibitem[Dai \latin{et~al.}(2014)Dai, Fei, Ma, Rodin, Wagner, McLeod, Liu, Gannett, Regan, Watanabe, Taniguchi, Thiemens, Dominguez, Neto, Zettl, Keilmann, Jarillo-Herrero, Fogler, and Basov]{PhPs_hBN}
Dai,~S. \latin{et~al.}  Tunable Phonon Polaritons in Atomically Thin van der Waals Crystals of Boron Nitride. \emph{Science} \textbf{2014}, \emph{343}, 1125--1129\relax
\mciteBstWouldAddEndPuncttrue
\mciteSetBstMidEndSepPunct{\mcitedefaultmidpunct}
{\mcitedefaultendpunct}{\mcitedefaultseppunct}\relax
\EndOfBibitem
\bibitem[Castilla \latin{et~al.}(2020)Castilla, Vangelidis, Pusapati, Goldstein, Autore, Slipchenko, Rajendran, Kim, Watanabe, Taniguchi, Martín-Moreno, Englund, Tielrooij, Hillenbrand, Lidorikis, and Koppens]{castilla_plasmonic_2020}
Castilla,~S. \latin{et~al.}  Plasmonic antenna coupling to hyperbolic phonon-polaritons for sensitive and fast mid-infrared photodetection with graphene. \emph{Nature Communications} \textbf{2020}, \emph{11}, 4872, Number: 1 Publisher: Nature Publishing Group\relax
\mciteBstWouldAddEndPuncttrue
\mciteSetBstMidEndSepPunct{\mcitedefaultmidpunct}
{\mcitedefaultendpunct}{\mcitedefaultseppunct}\relax
\EndOfBibitem
\bibitem[Duan \latin{et~al.}(2022)Duan, Alfaro-Mozaz, Taboada-Gutiérrez, Dolado, Álvarez Pérez, Titova, Bylinkin, Tresguerres-Mata, Martín-Sánchez, Liu, Edgar, Bandurin, Jarillo-Herrero, Hillenbrand, Nikitin, and Alonso-González]{PhPs_hBN2}
Duan,~J. \latin{et~al.}  Active and Passive Tuning of Ultranarrow Resonances in Polaritonic Nanoantennas. \emph{Advanced Materials} \textbf{2022}, \emph{34}, 2104954\relax
\mciteBstWouldAddEndPuncttrue
\mciteSetBstMidEndSepPunct{\mcitedefaultmidpunct}
{\mcitedefaultendpunct}{\mcitedefaultseppunct}\relax
\EndOfBibitem
\bibitem[Massicotte \latin{et~al.}(2016)Massicotte, Schmidt, Vialla, Schädler, Reserbat-Plantey, Watanabe, Taniguchi, Tielrooij, and Koppens]{massicotte_picosecond_2016}
Massicotte,~M.; Schmidt,~P.; Vialla,~F.; Schädler,~K.~G.; Reserbat-Plantey,~A.; Watanabe,~K.; Taniguchi,~T.; Tielrooij,~K.~J.; Koppens,~F. H.~L. Picosecond photoresponse in van der {Waals} heterostructures. \emph{Nature Nanotechnology} \textbf{2016}, \emph{11}, 42--46, 242\relax
\mciteBstWouldAddEndPuncttrue
\mciteSetBstMidEndSepPunct{\mcitedefaultmidpunct}
{\mcitedefaultendpunct}{\mcitedefaultseppunct}\relax
\EndOfBibitem
\bibitem[Gao \latin{et~al.}(2019)Gao, Zhou, Tsang, and Shu]{Fast_WG_integrated_tunnel_detector}
Gao,~Y.; Zhou,~G.; Tsang,~H.~K.; Shu,~C. High-speed van der Waals heterostructure tunneling photodiodes integrated on silicon nitride waveguides. \emph{Optica} \textbf{2019}, \emph{6}, 514--517\relax
\mciteBstWouldAddEndPuncttrue
\mciteSetBstMidEndSepPunct{\mcitedefaultmidpunct}
{\mcitedefaultendpunct}{\mcitedefaultseppunct}\relax
\EndOfBibitem
\bibitem[Schneider and Liu(2007)Schneider, and Liu]{QWIPs}
Schneider,~H.; Liu,~H.~C. \emph{Quantum well infrared photodetectors}; Springer, 2007\relax
\mciteBstWouldAddEndPuncttrue
\mciteSetBstMidEndSepPunct{\mcitedefaultmidpunct}
{\mcitedefaultendpunct}{\mcitedefaultseppunct}\relax
\EndOfBibitem
\bibitem[Ershov \latin{et~al.}(1995)Ershov, Ryzhii, and Hamaguchi]{QWIP_charge_buildup}
Ershov,~M.; Ryzhii,~V.; Hamaguchi,~C. {Contact and distributed effects in quantum well infrared photodetectors}. \emph{Applied Physics Letters} \textbf{1995}, \emph{67}, 3147--3149\relax
\mciteBstWouldAddEndPuncttrue
\mciteSetBstMidEndSepPunct{\mcitedefaultmidpunct}
{\mcitedefaultendpunct}{\mcitedefaultseppunct}\relax
\EndOfBibitem
\bibitem[Ryzhii \latin{et~al.}(2017)Ryzhii, Ryzhii, Svintsov, Leiman, Mitin, Shur, and Otsuji]{Ryzhii_GLIPs}
Ryzhii,~V.; Ryzhii,~M.; Svintsov,~D.; Leiman,~V.; Mitin,~V.; Shur,~M.~S.; Otsuji,~T. {Infrared photodetectors based on graphene van der Waals heterostructures}. \emph{Infrared Physics and Technology} \textbf{2017}, \emph{84}, 72--81\relax
\mciteBstWouldAddEndPuncttrue
\mciteSetBstMidEndSepPunct{\mcitedefaultmidpunct}
{\mcitedefaultendpunct}{\mcitedefaultseppunct}\relax
\EndOfBibitem
\bibitem[Liu \latin{et~al.}(2017)Liu, Rahman, Jiang, Li, and Fay]{Liu_tunnel_nonlinearity}
Liu,~L.; Rahman,~S.~M.; Jiang,~Z.; Li,~W.; Fay,~P. {Advanced Terahertz Sensing and Imaging Systems Based on Integrated III-V Interband Tunneling Devices}. \emph{Proceedings of the IEEE} \textbf{2017}, \emph{105}, 1020--1034\relax
\mciteBstWouldAddEndPuncttrue
\mciteSetBstMidEndSepPunct{\mcitedefaultmidpunct}
{\mcitedefaultendpunct}{\mcitedefaultseppunct}\relax
\EndOfBibitem
\bibitem[Gayduchenko \latin{et~al.}(2021)Gayduchenko, Xu, Alymov, Moskotin, Tretyakov, Taniguchi, Watanabe, Goltsman, Geim, Fedorov, Svintsov, and Bandurin]{gayduchenko_tunnel_2021}
Gayduchenko,~I.; Xu,~S.~G.; Alymov,~G.; Moskotin,~M.; Tretyakov,~I.; Taniguchi,~T.; Watanabe,~K.; Goltsman,~G.; Geim,~A.~K.; Fedorov,~G.; Svintsov,~D.; Bandurin,~D.~A. Tunnel field-effect transistors for sensitive terahertz detection. \emph{Nature Communications} \textbf{2021}, \emph{12}, 543, 30\relax
\mciteBstWouldAddEndPuncttrue
\mciteSetBstMidEndSepPunct{\mcitedefaultmidpunct}
{\mcitedefaultendpunct}{\mcitedefaultseppunct}\relax
\EndOfBibitem
\bibitem[Kazarinov and Suris(1972)Kazarinov, and Suris]{superlattice_Kazarinov}
Kazarinov,~R.; Suris,~R. Electric and electromagnetic properties of semiconductors with a superlattice. \emph{Sov. Phys. Semicond} \textbf{1972}, \emph{6}, 120--131\relax
\mciteBstWouldAddEndPuncttrue
\mciteSetBstMidEndSepPunct{\mcitedefaultmidpunct}
{\mcitedefaultendpunct}{\mcitedefaultseppunct}\relax
\EndOfBibitem
\bibitem[Rogalski \latin{et~al.}(2017)Rogalski, Martyniuk, and Kopytko]{Rogalski2017}
Rogalski,~A.; Martyniuk,~P.; Kopytko,~M. {InAs/GaSb type-II superlattice infrared detectors: Future prospect}. \emph{Applied Physics Reviews} \textbf{2017}, \emph{4}\relax
\mciteBstWouldAddEndPuncttrue
\mciteSetBstMidEndSepPunct{\mcitedefaultmidpunct}
{\mcitedefaultendpunct}{\mcitedefaultseppunct}\relax
\EndOfBibitem
\bibitem[Wei \latin{et~al.}(2005)Wei, Hood, Yau, Gin, Razeghi, Tidrow, and Nathan]{Quantum_cascade_detector}
Wei,~Y.; Hood,~A.; Yau,~H.; Gin,~A.; Razeghi,~M.; Tidrow,~M.~Z.; Nathan,~V. Uncooled operation of type-II InAs/GaSb superlattice photodiodes in the midwavelength infrared range. \emph{Applied Physics Letters} \textbf{2005}, \emph{86}, 1--3\relax
\mciteBstWouldAddEndPuncttrue
\mciteSetBstMidEndSepPunct{\mcitedefaultmidpunct}
{\mcitedefaultendpunct}{\mcitedefaultseppunct}\relax
\EndOfBibitem
\bibitem[Faist \latin{et~al.}(1994)Faist, Capasso, Sivco, Sirtori, Hutchinson, and Cho]{QCL_Faist}
Faist,~J.; Capasso,~F.; Sivco,~D.~L.; Sirtori,~C.; Hutchinson,~A.~L.; Cho,~A.~Y. {Quantum Cascade Laser}. \emph{Science} \textbf{1994}, \emph{264}, 553--556\relax
\mciteBstWouldAddEndPuncttrue
\mciteSetBstMidEndSepPunct{\mcitedefaultmidpunct}
{\mcitedefaultendpunct}{\mcitedefaultseppunct}\relax
\EndOfBibitem
\bibitem[Ryzhii \latin{et~al.}(2014)Ryzhii, Otsuji, Aleshkin, Dubinov, Ryzhii, Mitin, and Shur]{ryzhii_voltage-tunable_2014}
Ryzhii,~V.; Otsuji,~T.; Aleshkin,~V.~Y.; Dubinov,~A.~A.; Ryzhii,~M.; Mitin,~V.; Shur,~M.~S. Voltage-tunable terahertz and infrared photodetectors based on double-graphene-layer structures. \emph{Applied Physics Letters} \textbf{2014}, \emph{104}, 163505, 31\relax
\mciteBstWouldAddEndPuncttrue
\mciteSetBstMidEndSepPunct{\mcitedefaultmidpunct}
{\mcitedefaultendpunct}{\mcitedefaultseppunct}\relax
\EndOfBibitem
\bibitem[Ryzhii \latin{et~al.}(2013)Ryzhii, Dubinov, Aleshkin, Ryzhii, and Otsuji]{Ryzhii_tunnel_injection_laser}
Ryzhii,~V.; Dubinov,~A.~A.; Aleshkin,~V.~Y.; Ryzhii,~M.; Otsuji,~T. {Injection terahertz laser using the resonant inter-layer radiative transitions in double-graphene-layer structure}. \emph{Applied Physics Letters} \textbf{2013}, \emph{103}, 10--14\relax
\mciteBstWouldAddEndPuncttrue
\mciteSetBstMidEndSepPunct{\mcitedefaultmidpunct}
{\mcitedefaultendpunct}{\mcitedefaultseppunct}\relax
\EndOfBibitem
\bibitem[Xie \latin{et~al.}(2023)Xie, Ji, Wu, Zhang, Jin, Watanabe, Taniguchi, Liu, and Cai]{xie_probing_2023}
Xie,~B.; Ji,~Z.; Wu,~J.; Zhang,~R.; Jin,~Y.; Watanabe,~K.; Taniguchi,~T.; Liu,~Z.; Cai,~X. Probing the {Inelastic} {Electron} {Tunneling} via the {Photocurrent} in a {Vertical} {Graphene} van der {Waals} {Heterostructure}. \emph{ACS Nano} \textbf{2023}, Publisher: American Chemical Society\relax
\mciteBstWouldAddEndPuncttrue
\mciteSetBstMidEndSepPunct{\mcitedefaultmidpunct}
{\mcitedefaultendpunct}{\mcitedefaultseppunct}\relax
\EndOfBibitem
\bibitem[Ma \latin{et~al.}(2016)Ma, Andersen, Nair, Gabor, Massicotte, Lui, Young, Fang, Watanabe, Taniguchi, Kong, Gedik, Koppens, and Jarillo-Herrero]{ma_tuning_2016}
Ma,~Q.; Andersen,~T.~I.; Nair,~N.~L.; Gabor,~N.~M.; Massicotte,~M.; Lui,~C.~H.; Young,~A.~F.; Fang,~W.; Watanabe,~K.; Taniguchi,~T.; Kong,~J.; Gedik,~N.; Koppens,~F. H.~L.; Jarillo-Herrero,~P. Tuning ultrafast electron thermalization pathways in a van der {Waals} heterostructure. \emph{Nature Physics} \textbf{2016}, \emph{12}, 455--459, 103\relax
\mciteBstWouldAddEndPuncttrue
\mciteSetBstMidEndSepPunct{\mcitedefaultmidpunct}
{\mcitedefaultendpunct}{\mcitedefaultseppunct}\relax
\EndOfBibitem
\bibitem[Kuzmina \latin{et~al.}(2021)Kuzmina, Parzefall, Back, Taniguchi, Watanabe, Jain, and Novotny]{kuzmina_resonant_2021}
Kuzmina,~A.; Parzefall,~M.; Back,~P.; Taniguchi,~T.; Watanabe,~K.; Jain,~A.; Novotny,~L. Resonant {Light} {Emission} from {Graphene}/{Hexagonal} {Boron} {Nitride}/{Graphene} {Tunnel} {Junctions}. \emph{Nano Letters} \textbf{2021}, \emph{21}, 8332--8339, 3\relax
\mciteBstWouldAddEndPuncttrue
\mciteSetBstMidEndSepPunct{\mcitedefaultmidpunct}
{\mcitedefaultendpunct}{\mcitedefaultseppunct}\relax
\EndOfBibitem
\bibitem[Yadav \latin{et~al.}(2016)Yadav, Tombet, Watanabe, Arnold, Ryzhii, and Otsuji]{yadav_terahertz_2016}
Yadav,~D.; Tombet,~S.~B.; Watanabe,~T.; Arnold,~S.; Ryzhii,~V.; Otsuji,~T. Terahertz wave generation and detection in double-graphene layered van der {Waals} heterostructures. \emph{2D Materials} \textbf{2016}, \emph{3}, 045009, 51\relax
\mciteBstWouldAddEndPuncttrue
\mciteSetBstMidEndSepPunct{\mcitedefaultmidpunct}
{\mcitedefaultendpunct}{\mcitedefaultseppunct}\relax
\EndOfBibitem
\bibitem[Mishchenko \latin{et~al.}(2014)Mishchenko, Tu, Cao, Gorbachev, Wallbank, Greenaway, Morozov, Morozov, Zhu, Wong, Withers, Woods, Kim, Watanabe, Taniguchi, Vdovin, Makarovsky, Fromhold, Fal'ko, Geim, Eaves, and Novoselov]{mishchenko_twist-controlled_2014}
Mishchenko,~A. \latin{et~al.}  Twist-controlled resonant tunnelling in graphene/boron nitride/graphene heterostructures. \emph{Nature Nanotechnology} \textbf{2014}, \emph{9}, 808--813, 355\relax
\mciteBstWouldAddEndPuncttrue
\mciteSetBstMidEndSepPunct{\mcitedefaultmidpunct}
{\mcitedefaultendpunct}{\mcitedefaultseppunct}\relax
\EndOfBibitem
\bibitem[Ghazaryan \latin{et~al.}(2021)Ghazaryan, Misra, Vdovin, Watanabe, Taniguchi, Morozov, Mishchenko, and Novoselov]{ghazaryan_twisted_2021}
Ghazaryan,~D.~A.; Misra,~A.; Vdovin,~E.~E.; Watanabe,~K.; Taniguchi,~T.; Morozov,~S.~V.; Mishchenko,~A.; Novoselov,~K.~S. Twisted monolayer and bilayer graphene for vertical tunneling transistors. \emph{Applied Physics Letters} \textbf{2021}, \emph{118}, 183106\relax
\mciteBstWouldAddEndPuncttrue
\mciteSetBstMidEndSepPunct{\mcitedefaultmidpunct}
{\mcitedefaultendpunct}{\mcitedefaultseppunct}\relax
\EndOfBibitem
\bibitem[Chandni \latin{et~al.}(2016)Chandni, Watanabe, Taniguchi, and Eisenstein]{chandni_signatures_2016}
Chandni,~U.; Watanabe,~K.; Taniguchi,~T.; Eisenstein,~J.~P. Signatures of {Phonon} and {Defect}-{Assisted} {Tunneling} in {Planar} {Metal}–{Hexagonal} {Boron} {Nitride}–{Graphene} {Junctions}. \emph{Nano Letters} \textbf{2016}, \emph{16}, 7982--7987, 45\relax
\mciteBstWouldAddEndPuncttrue
\mciteSetBstMidEndSepPunct{\mcitedefaultmidpunct}
{\mcitedefaultendpunct}{\mcitedefaultseppunct}\relax
\EndOfBibitem
\bibitem[Greenaway \latin{et~al.}(2018)Greenaway, Vdovin, Ghazaryan, Misra, Mishchenko, Cao, Wang, Wallbank, Holwill, Khanin, Morozov, Watanabe, Taniguchi, Makarovsky, Fromhold, Patanè, Geim, Fal’ko, Novoselov, and Eaves]{greenaway_tunnel_2018}
Greenaway,~M.~T. \latin{et~al.}  Tunnel spectroscopy of localised electronic states in hexagonal boron nitride. \emph{Communications Physics} \textbf{2018}, \emph{1}, 1--7, 27\relax
\mciteBstWouldAddEndPuncttrue
\mciteSetBstMidEndSepPunct{\mcitedefaultmidpunct}
{\mcitedefaultendpunct}{\mcitedefaultseppunct}\relax
\EndOfBibitem
\bibitem[Crossno \latin{et~al.}(2015)Crossno, Liu, Ohki, Kim, and Fong]{crossno_development_2015}
Crossno,~J.; Liu,~X.; Ohki,~T.~A.; Kim,~P.; Fong,~K.~C. Development of high frequency and wide bandwidth {Johnson} noise thermometry. \emph{Applied Physics Letters} \textbf{2015}, \emph{106}, 023121, 30\relax
\mciteBstWouldAddEndPuncttrue
\mciteSetBstMidEndSepPunct{\mcitedefaultmidpunct}
{\mcitedefaultendpunct}{\mcitedefaultseppunct}\relax
\EndOfBibitem
\bibitem[Fong and Schwab(2012)Fong, and Schwab]{fong_ultrasensitive_2012}
Fong,~K.~C.; Schwab,~K.~C. Ultrasensitive and {Wide}-{Bandwidth} {Thermal} {Measurements} of {Graphene} at {Low} {Temperatures}. \emph{Physical Review X} \textbf{2012}, \emph{2}, 031006, Publisher: American Physical Society\relax
\mciteBstWouldAddEndPuncttrue
\mciteSetBstMidEndSepPunct{\mcitedefaultmidpunct}
{\mcitedefaultendpunct}{\mcitedefaultseppunct}\relax
\EndOfBibitem
\bibitem[Fong \latin{et~al.}(2013)Fong, Wollman, Ravi, Chen, Clerk, Shaw, Leduc, and Schwab]{fong_measurement_2013}
Fong,~K.~C.; Wollman,~E.~E.; Ravi,~H.; Chen,~W.; Clerk,~A.~A.; Shaw,~M.~D.; Leduc,~H.~G.; Schwab,~K.~C. Measurement of the {Electronic} {Thermal} {Conductance} {Channels} and {Heat} {Capacity} of {Graphene} at {Low} {Temperature}. \emph{Physical Review X} \textbf{2013}, \emph{3}, 041008, 61\relax
\mciteBstWouldAddEndPuncttrue
\mciteSetBstMidEndSepPunct{\mcitedefaultmidpunct}
{\mcitedefaultendpunct}{\mcitedefaultseppunct}\relax
\EndOfBibitem
\bibitem[Betz \latin{et~al.}(2013)Betz, Jhang, Pallecchi, Ferreira, Fève, Berroir, and Plaçais]{betz_supercollision_2013}
Betz,~A.~C.; Jhang,~S.~H.; Pallecchi,~E.; Ferreira,~R.; Fève,~G.; Berroir,~J.-M.; Plaçais,~B. Supercollision cooling in undoped graphene. \emph{Nature Physics} \textbf{2013}, \emph{9}, 109--112, 171\relax
\mciteBstWouldAddEndPuncttrue
\mciteSetBstMidEndSepPunct{\mcitedefaultmidpunct}
{\mcitedefaultendpunct}{\mcitedefaultseppunct}\relax
\EndOfBibitem
\bibitem[Tikhonov \latin{et~al.}(2016)Tikhonov, Shovkun, Ercolani, Rossella, Rocci, Sorba, Roddaro, and Khrapai]{tikhonov_noise_2016}
Tikhonov,~E.~S.; Shovkun,~D.~V.; Ercolani,~D.; Rossella,~F.; Rocci,~M.; Sorba,~L.; Roddaro,~S.; Khrapai,~V.~S. Noise thermometry applied to thermoelectric measurements in {InAs} nanowires. \emph{Semiconductor Science and Technology} \textbf{2016}, \emph{31}, 104001, 16\relax
\mciteBstWouldAddEndPuncttrue
\mciteSetBstMidEndSepPunct{\mcitedefaultmidpunct}
{\mcitedefaultendpunct}{\mcitedefaultseppunct}\relax
\EndOfBibitem
\bibitem[Kretinin \latin{et~al.}(2014)Kretinin, Cao, Tu, Yu, Jalil, Novoselov, Haigh, Gholinia, Mishchenko, Lozada, Georgiou, Woods, Withers, Blake, Eda, Wirsig, Hucho, Watanabe, Taniguchi, Geim, and Gorbachev]{kretinin_electronic_2014}
Kretinin,~A.~V. \latin{et~al.}  Electronic {Properties} of {Graphene} {Encapsulated} with {Different} {Two}-{Dimensional} {Atomic} {Crystals}. \emph{Nano Letters} \textbf{2014}, \emph{14}, 3270--3276, Publisher: American Chemical Society\relax
\mciteBstWouldAddEndPuncttrue
\mciteSetBstMidEndSepPunct{\mcitedefaultmidpunct}
{\mcitedefaultendpunct}{\mcitedefaultseppunct}\relax
\EndOfBibitem
\bibitem[Vdovin \latin{et~al.}(2016)Vdovin, Mishchenko, Greenaway, Zhu, Ghazaryan, Misra, Cao, Morozov, Makarovsky, Fromhold, Patanè, Slotman, Katsnelson, Geim, Novoselov, and Eaves]{vdovin_phonon-assisted_2016}
Vdovin,~E. \latin{et~al.}  Phonon-{Assisted} {Resonant} {Tunneling} of {Electrons} in {Graphene}--{Boron} {Nitride} {Transistors}. \emph{Physical Review Letters} \textbf{2016}, \emph{116}, 186603, 74\relax
\mciteBstWouldAddEndPuncttrue
\mciteSetBstMidEndSepPunct{\mcitedefaultmidpunct}
{\mcitedefaultendpunct}{\mcitedefaultseppunct}\relax
\EndOfBibitem
\bibitem[Fainberg(2013)]{fainberg2013photon}
Fainberg,~B.~D. Photon-assisted tunneling through molecular conduction junctions with graphene electrodes. \emph{Physical Review B} \textbf{2013}, \emph{88}, 245435\relax
\mciteBstWouldAddEndPuncttrue
\mciteSetBstMidEndSepPunct{\mcitedefaultmidpunct}
{\mcitedefaultendpunct}{\mcitedefaultseppunct}\relax
\EndOfBibitem
\bibitem[Platero and Aguado(2004)Platero, and Aguado]{platero2004photon}
Platero,~G.; Aguado,~R. Photon-assisted transport in semiconductor nanostructures. \emph{Physics Reports} \textbf{2004}, \emph{395}, 1--157\relax
\mciteBstWouldAddEndPuncttrue
\mciteSetBstMidEndSepPunct{\mcitedefaultmidpunct}
{\mcitedefaultendpunct}{\mcitedefaultseppunct}\relax
\EndOfBibitem
\bibitem[Kleinekath{\"o}fer \latin{et~al.}(2006)Kleinekath{\"o}fer, Li, Welack, and Schreiber]{kleinekathofer2006switching}
Kleinekath{\"o}fer,~U.; Li,~G.; Welack,~S.; Schreiber,~M. Switching the current through model molecular wires with Gaussian laser pulses. \emph{Europhysics Letters} \textbf{2006}, \emph{75}, 139\relax
\mciteBstWouldAddEndPuncttrue
\mciteSetBstMidEndSepPunct{\mcitedefaultmidpunct}
{\mcitedefaultendpunct}{\mcitedefaultseppunct}\relax
\EndOfBibitem
\bibitem[Tien and Gordon(1963)Tien, and Gordon]{tien1963multiphoton}
Tien,~P.; Gordon,~J. Multiphoton process observed in the interaction of microwave fields with the tunneling between superconductor films. \emph{Physical Review} \textbf{1963}, \emph{129}, 647\relax
\mciteBstWouldAddEndPuncttrue
\mciteSetBstMidEndSepPunct{\mcitedefaultmidpunct}
{\mcitedefaultendpunct}{\mcitedefaultseppunct}\relax
\EndOfBibitem
\bibitem[Georgiou \latin{et~al.}(2013)Georgiou, Jalil, Belle, Britnell, Gorbachev, Morozov, Kim, Gholinia, Haigh, Makarovsky, Eaves, Ponomarenko, Geim, Novoselov, and Mishchenko]{georgiou_vertical_2013}
Georgiou,~T.; Jalil,~R.; Belle,~B.~D.; Britnell,~L.; Gorbachev,~R.~V.; Morozov,~S.~V.; Kim,~Y.-J.; Gholinia,~A.; Haigh,~S.~J.; Makarovsky,~O.; Eaves,~L.; Ponomarenko,~L.~A.; Geim,~A.~K.; Novoselov,~K.~S.; Mishchenko,~A. Vertical field-effect transistor based on graphene–{WS2} heterostructures for flexible and transparent electronics. \emph{Nature Nanotechnology} \textbf{2013}, \emph{8}, 100--103, Number: 2 Publisher: Nature Publishing Group\relax
\mciteBstWouldAddEndPuncttrue
\mciteSetBstMidEndSepPunct{\mcitedefaultmidpunct}
{\mcitedefaultendpunct}{\mcitedefaultseppunct}\relax
\EndOfBibitem
\bibitem[Bai \latin{et~al.}(2022)Bai, Xiao, Luo, Li, Peng, Zhu, Luo, Zhu, Qin, and Novoselov]{bai_highly_2022}
Bai,~Z.; Xiao,~Y.; Luo,~Q.; Li,~M.; Peng,~G.; Zhu,~Z.; Luo,~F.; Zhu,~M.; Qin,~S.; Novoselov,~K. Highly {Tunable} {Carrier} {Tunneling} in {Vertical} {Graphene}–{WS2}–{Graphene} van der {Waals} {Heterostructures}. \emph{ACS Nano} \textbf{2022}, \emph{16}, 7880--7889, 1\relax
\mciteBstWouldAddEndPuncttrue
\mciteSetBstMidEndSepPunct{\mcitedefaultmidpunct}
{\mcitedefaultendpunct}{\mcitedefaultseppunct}\relax
\EndOfBibitem
\end{mcitethebibliography}


\providecommand{\latin}[1]{#1}
\makeatletter
\providecommand{\doi}
  {\begingroup\let\do\@makeother\dospecials
  \catcode`\{=1 \catcode`\}=2 \doi@aux}
\providecommand{\doi@aux}[1]{\endgroup\texttt{#1}}
\makeatother
\providecommand*\mcitethebibliography{\thebibliography}
\csname @ifundefined\endcsname{endmcitethebibliography}  {\let\endmcitethebibliography\endthebibliography}{}
\begin{mcitethebibliography}{2}
\providecommand*\natexlab[1]{#1}
\providecommand*\mciteSetBstSublistMode[1]{}
\providecommand*\mciteSetBstMaxWidthForm[2]{}
\providecommand*\mciteBstWouldAddEndPuncttrue
  {\def\EndOfBibitem{\unskip.}}
\providecommand*\mciteBstWouldAddEndPunctfalse
  {\let\EndOfBibitem\relax}
\providecommand*\mciteSetBstMidEndSepPunct[3]{}
\providecommand*\mciteSetBstSublistLabelBeginEnd[3]{}
\providecommand*\EndOfBibitem{}
\mciteSetBstSublistMode{f}
\mciteSetBstMaxWidthForm{subitem}{(\alph{mcitesubitemcount})}
\mciteSetBstSublistLabelBeginEnd
  {\mcitemaxwidthsubitemform\space}
  {\relax}
  {\relax}

\bibitem[Greenaway \latin{et~al.}(2018)Greenaway, Vdovin, Ghazaryan, Misra, Mishchenko, Cao, Wang, Wallbank, Holwill, Khanin, Morozov, Watanabe, Taniguchi, Makarovsky, Fromhold, Patanè, Geim, Fal’ko, Novoselov, and Eaves]{greenaway_tunnel_2018}
Greenaway,~M.~T. \latin{et~al.}  Tunnel spectroscopy of localised electronic states in hexagonal boron nitride. \emph{Communications Physics} \textbf{2018}, \emph{1}, 1--7, 27\relax
\mciteBstWouldAddEndPuncttrue
\mciteSetBstMidEndSepPunct{\mcitedefaultmidpunct}
{\mcitedefaultendpunct}{\mcitedefaultseppunct}\relax
\EndOfBibitem
\end{mcitethebibliography}


\end{document}


\maketitle

\section{I. Measurements details}

\begin{figure*}
    \includegraphics[width=0.8\textwidth]{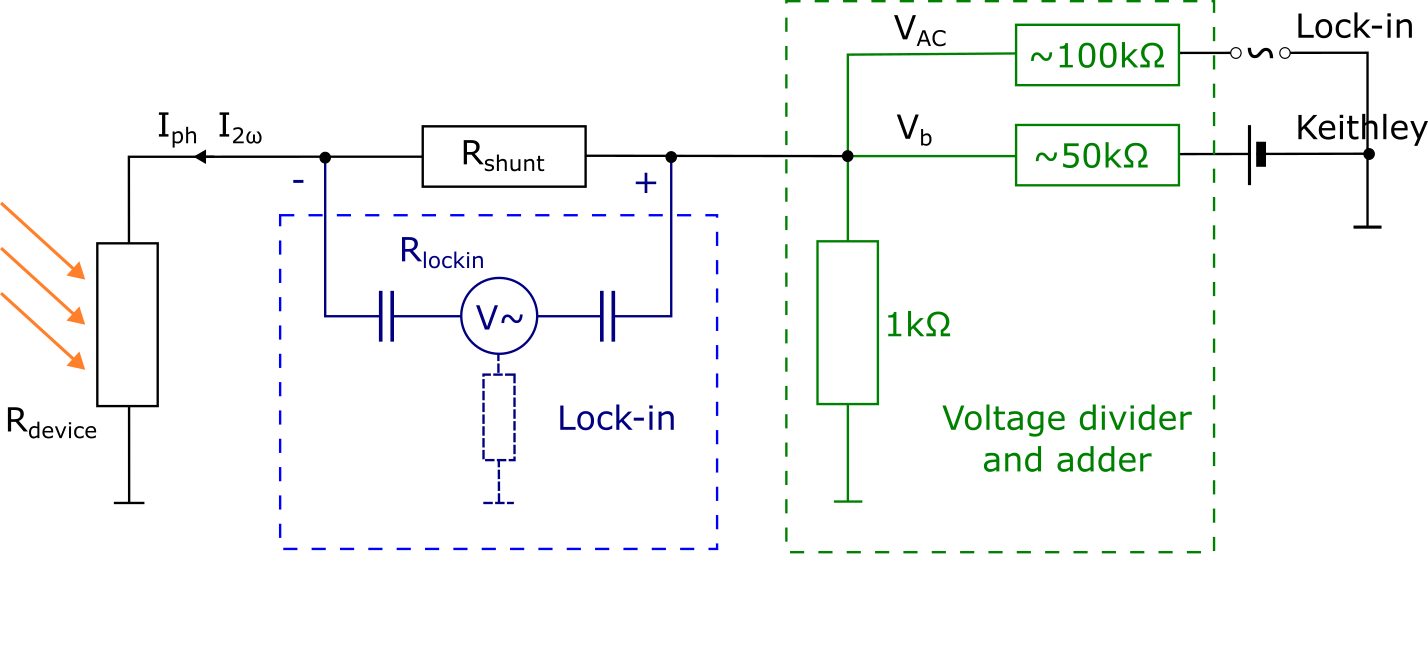}
    \caption{Electrical scheme used for photocurrent and $\dII$ measurements. The voltage divider is highlighted in green, and the simplified equivalent scheme of the lock-in amplifier is shown in blue.}
    \label{fig}
\end{figure*}

The electrical circuit for measuring photocurrent and $\dII$ is shown in Figure S1. For $\dII$ measurements the Keithley DC voltage and lock-in SR860 output AC sine voltage was summed up by a voltage divider with ratios 1:53 and 1:104, respectively. This allowed us to apply a DC bias to the sample in the range of $-0.4$ to $0.4$~V with a small AC component of $V_{\rm AC}=4.8$~mV (rms). For photocurrent measurements, $V_{\rm AC}=0$.

Due to a short circuit between the bottom graphene metallic contact and the ground (likely occurring during bonding), the contact was not available for measuring current. Therefore, we calculated the current through the sample based on voltage measurements using a shunt resistance of $R_{shunt}=1.3$~M$\Omega$. The lock-in amplifier has an input impedance of $R_{lockin}=10$~M$\Omega$ and only passes AC current. It also has an internal resistance to the ground of about $10$~M$\Omega$. Taking this into account, the voltage divider and shunt resistances were carefully chosen to ensure they did not influence the measurements: $1$~k$\Omega \ll R_{shunt} \ll R_{device}$, $R_{shunt} \ll R_{lockin}$.

To carefully capture the phase of the photocurrent during synchronous detection, a low frequency of 8 Hz was chosen, where 8~Hz~$\ll 1/RC$. The high $RC$ value of the sample arises from its high resistance and capacitance of the circuit.

\begin{figure*}
    \includegraphics[width=\textwidth]{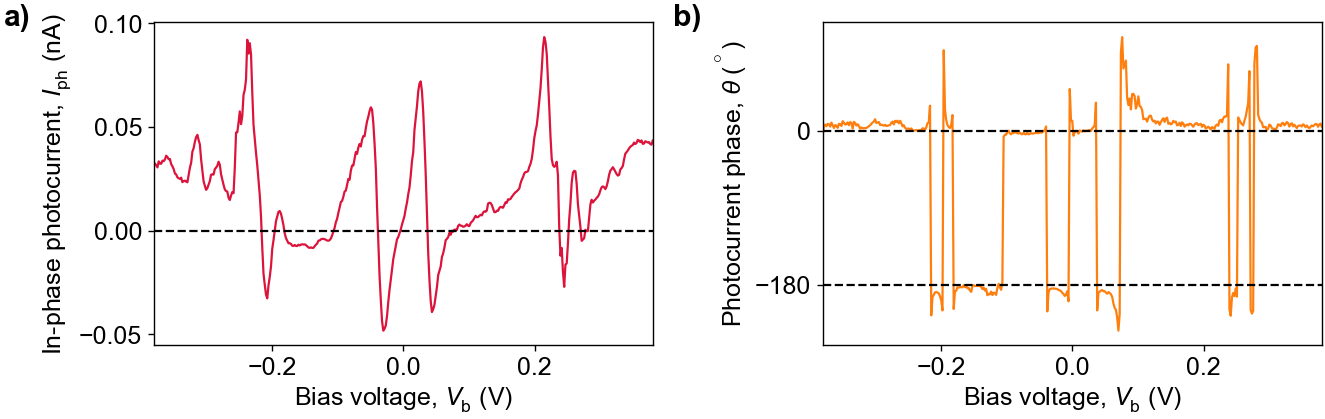}
    \caption{Photocurrent at $\Vg=-1$~V after binding the chopper phase: (a) in-phase component, measured with lock-in and (b) phase. Being always around 0 and 180 degrees, it indirectly indicates that 1) the chopper phase is bound correctly, 2) parasitic capacitances don't influence on measurements.}
    \label{fig}
\end{figure*}

Collimated laser $D=25$~mm beam was focused using $F=50$~mm aspheric ZnSe lens.  Scanning was done by moving focusing lens held on the XY-stage. For used small scanning area size lens decentering does not brings aberrations but does displace IR beam.

Binding of chopper phase was done by comparing photocurrent phase with that obtained in case of laser current modulation and additionally controlled from amplified waveforms on the oscilloscope. The resulting photocurrent in-phase component and phase are shown in Figure S2.

\begin{figure*}
    \includegraphics[width=0.5\textwidth]{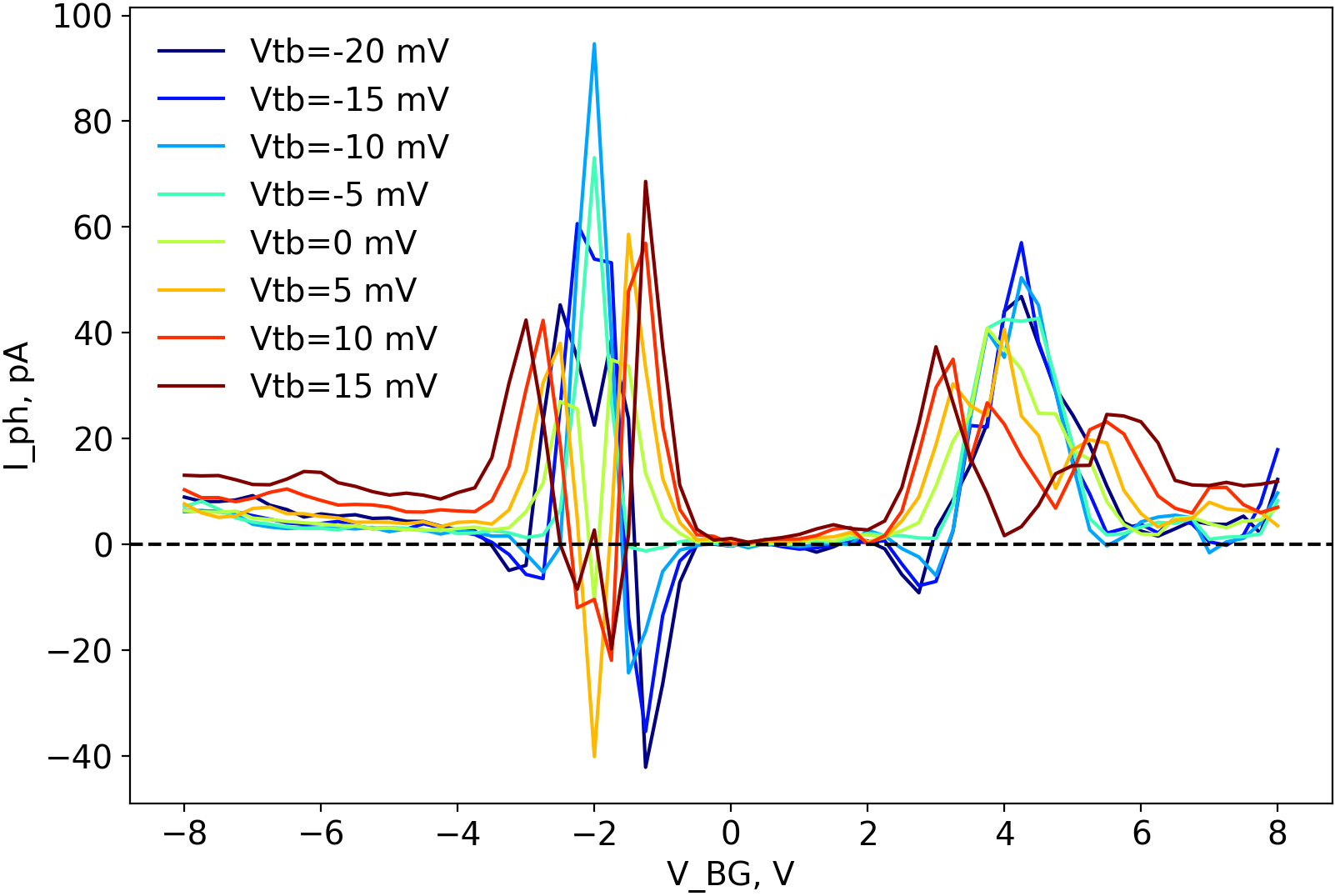}
    \caption{Slices of photocurrent map from Figure 2d at bias voltages near zero.}
    \label{fig}
\end{figure*}

\section{II. Photocurrent measurements of the device \#2}

\begin{figure*}
    \includegraphics[width=1\textwidth]{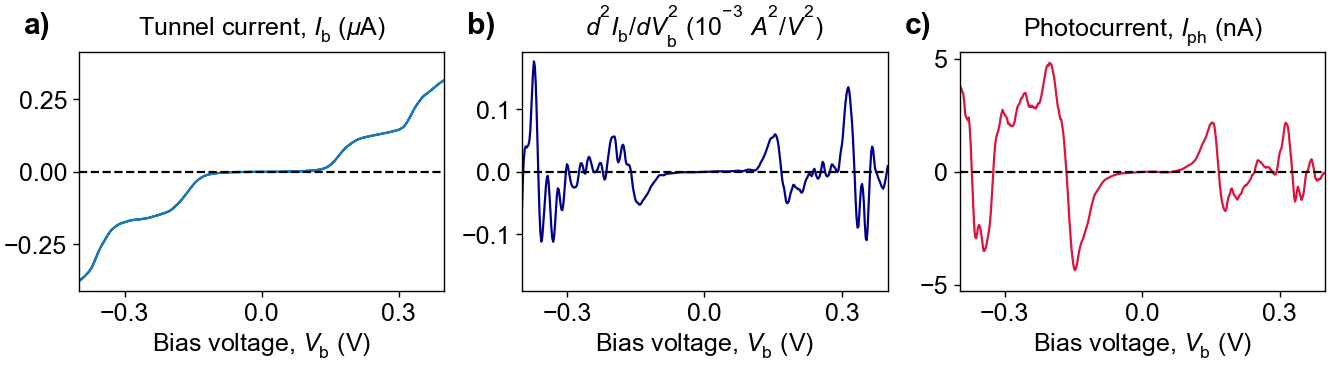}
    \caption{Transport and photo measurements of device \#2 at $T=9$~K for $\Vg=10$~V. (a) \IV{} characteristics. (b) $\dII$. (c) Photocurrent under $\lambda=6.0$~$\mu$m illumination.}
    \label{fig}
\end{figure*}

Device \#2 shows same features in transport and photocurrent measurements as device \#1 presented in the main text. The photocurrent is proportional to second derivative of $\IonV$ and maximizes at bias values where impurity-assisted tunneling occurs. Having a larger tunnel area, device demonstrates 10 times large current and 100 times larger photocurrent (Figure S3). This Gr/hBN/Gr hBN-encapsulated stack with 1~nm barrier was made of multilayer graphene (2L on top and 3L on the bottom) and has silicon gate separated from device by 70~nm of bottom hBN and 280~nm SiO$_2$.

\section{III. Theory of photocurrent generation}
\subsection*{Electrostatic model}
An electrostatic model of a device (Figure 1a) with a tunnel barrier width $d$ and a distance to the gate $d_{\rm g}$, assumes that the structure is infinite along the xy plane and is given by

\begin{equation}
\begin{aligned}
e V_{\mathrm{b}} & =\mu_{\mathrm{b}}-\mu_{\mathrm{t}}-e d F_{\mathrm{b}} \\
e V_{\mathrm{gate}} & =\mu_{\mathrm{b}}+e d_{\mathrm{g}} F_{\mathrm{g}}
\end{aligned}
\label{ElectStat}
\end{equation}

where $F_{\mathrm{g}}$ is the electric field between the bottom graphene layer and the gate,  $F_{\mathrm{b}}$ is the electric field within the tunnel barrier region, $\mu_{\mathrm{b,t}}$ are the chemical potentials in the bottom and top graphene layers respectively measured with respect to the Dirac point. The equation \ref{ElectStat} was solved using Gauss law to find charges in graphene layers and on the gate surface. A similar approach was considered in \cite{greenaway_tunnel_2018}.

\subsection*{Tunnel current calculation}
The step in \IV{}-curve (and a spike in conductance) occurs as soon as the Fermi level in either of graphene layers crosses the level of impurity. The current resulting from tunneling through a localized state in the hBN is given by

\begin{equation}
I_{\rm b}= \frac{e}{\hbar} \int_{-\infty}^{\infty} d E \frac{\gamma_{\mathrm{b}} \gamma_{\mathrm{t}}}{\gamma_{\mathrm{b}}+\gamma_{\mathrm{t}}}\left(f_{\mathrm{t}}-f_{\mathrm{b}}\right) \Gamma\left(E-E_{\mathrm{i}}\right),
\label{TunnelCurrent}
\end{equation}

where $E_i$ is the energy of the impurity level, $\gamma_{\rm t,b}/\hbar$ is the electronic tunnelling rate from a localised state into the bottom and top layers respectively, $f_{\mathrm{t,b}}=1 /\left(1+\exp \left(\left(E-\mu_{\mathrm{t,b}}\right) / k_{\mathrm{B}} T_{\mathrm{t,b}}\right)\right)$ is the Fermi distribution function of the top and bottom layers of graphene with temperatures $T_{\rm t}$ and $T_{\rm b}$ respectively, $\Gamma(E)$ is a Gaussian function with a full width half maximum of $\gamma = \gamma_\mathrm{t}+\gamma_\mathrm{b}$. This is described in detail in \cite{greenaway_tunnel_2018}. To fit the experimental data, we considered two impurity levels with energies $E_{i1}=100$ meV and $E_{i2}=-70$~meV reckoned from the Dirac points of unbiased layers. The impurity \#1 is located in the middle of the barrier, while impurity \#2 is between the second and third hBN layers, counting from the bottom graphene. In the calculations we used $\gamma_1 = 4.25$ meV and $\gamma_2 = 2.5$ meV.


\subsection{Proportionality of $\dII$ and $d\Ib/dT$}

To simplify the calculations, we assume that in the equation (S2) $\Gamma(E)$ is a slowly changing compared to the distribution functions $f_{\rm{t,b}}$ and consider a linear approach $\Gamma(E-E_i) = \Gamma(\mu_{\rm{t,b}}-E_i)+(E-E_i)\Gamma'(\mu_{\rm{t,b}}-E_i)$, where $\Gamma'(E)$ is the derivative of $\Gamma(E)$. This approach is applicable with $\gamma_{1,2} \gg k_B T_0$. Neglecting the effects of quantum capacitance, it can be assumed that the Fermi levels depend linearly on the bias, which leads after integrating the equation (S2) to the following expressions:

\begin{equation}
    \frac{1}{k_{\rm B}} \frac{dI_{\rm{b}}}{dT_{\rm{t,b}}} = \frac{\pi^2}{3} \frac{e}{\hbar} \gamma_{\rm eff} k_{\rm B} T_{\rm t,b} \Gamma'(\mu_{\rm t,b}-E_i),
\end{equation}

\begin{equation}
 \frac{d^2I_{\rm{b}}}{dV_{\rm{b}}^2} = \frac{e}{2 \hbar} \gamma_{\rm eff} \left( \alpha_{\rm t} \Gamma'(\mu_{\rm t}-E_i) - \alpha_{\rm b} \Gamma'(\mu_{\rm b}-E_i) \right),
\end{equation}

where $\gamma_{\rm eff} = \gamma_{\rm t} \gamma_{\rm b}/ (\gamma_{\rm t}+\gamma_{\rm b})$ and $\alpha_{\rm t,b} = (d\mu_{\rm t,b}/ d V_{\rm b})^2 - 4 (d\mu_{\rm t,b}/ d V_{\rm b}) (d E_i / d V_{\rm b})$. Thus, we get the following relation between $\dII$ and $d\Ib/dT$:

\begin{equation}
 \frac{d^2I_{\rm{b}}}{dV_{\rm{b}}^2} = \frac{3}{2 \pi^2} \left( \frac{\alpha_{\rm t}}{k^2_B T_{\rm t}} \frac{dI_{\rm{b}}}{dT_{\rm{t}}}  - \frac{\alpha_{\rm b}}{k^2_B T_{\rm b}} \frac{dI_{\rm{b}}}{dT_{\rm{b}}} \right).
\end{equation}

\bibliography{bibtexa,Bibliography-extra}